\documentclass[pdflatex,sn-basic,referee]{sn-jnl}

\usepackage{graphicx}%
\usepackage{multirow}%
\usepackage{amsmath,amssymb,amsfonts}%
\usepackage{amsthm}%
\usepackage{mathrsfs}%
\usepackage[title]{appendix}%
\usepackage{xcolor}%
\usepackage{textcomp}%
\usepackage{manyfoot}%
\usepackage{booktabs}%
\usepackage{algorithm}%
\usepackage{algorithmicx}%
\usepackage{algpseudocode}%
\usepackage{listings}%

\theoremstyle{thmstyleone}%
\newtheorem{theorem}{Theorem}
\theoremstyle{thmstyletwo}%
\newtheorem{example}{Example}%
\newtheorem{remark}{Remark}[section]
\newtheorem{assumption}{Assumption}
\newtheorem{lemma}{Lemma}

\theoremstyle{thmstylethree}%

\begin{document}

\title[Fair Conformal Prediction for Incomplete Covariate Data]{Fair Conformal Prediction for Incomplete Covariate Data}


\author[1]{\fnm{Jingsen} \sur{Kong}}\email{jingsen0111@163.com}

\author[1]{\fnm{Yiming} \sur{Liu}}\email{liuyiming@jnu.edu.cn}

\author*[1]{\fnm{Guangren} \sur{Yang}}\email{tygr@jnu.edu.cn}

\affil*[1]{\orgdiv{Department of Statistics and Data Science}, \orgname{Jinan University}, \orgaddress{\city{Guangzhou}, \country{China}}}




\abstract{Conformal prediction provides a distribution-free framework for uncertainty quantification. This study explores the application of conformal prediction in scenarios where covariates are missing, which introduces significant challenges for uncertainty quantification. Building on the framework of nonexchangeable conformal prediction, we demonstrate that coverage guarantees depend on the mask. To address this, we propose a nonexchangeable conformal prediction method for missing covariates that satisfies both marginal and mask-conditional validity. However, as this method does not ensure asymptotic conditional validity, we further introduce a localized conformal prediction approach that employs a novel score function based on kernel smoothing. This method achieves marginal, mask-conditional, and asymptotic conditional validity under certain assumptions. Extensive simulation studies and real-data analysis demonstrate the advantages of these proposed methods.}




\maketitle

\section{Introduction}
Conformal prediction is a general framework for distribution-free uncertainty quantification \citep{vovk2005algorithmic, shafer2008tutorial, lei2018distribution, tibshirani2019conformal, angelopoulos2023conformal, barber2023conformal}. Given a random vector \( X \in \mathcal{X} \subseteq \mathbb{R}^d \), which contains \( d \)-dimensional explanatory variables, the goal of conformal prediction is to construct a prediction set \( \mathcal{C}_{\alpha}(\cdot) \) for the response variable \( Y \in \mathcal{Y} \), such that for a pre-specified miscoverage level \( \alpha \in (0, 1) \), the following coverage guarantee holds:
\begin{equation*}
	\mathbb{P}\left\{ Y \in \mathcal{C}_{\alpha}(X) \right\} \geq 1 - \alpha.
\end{equation*}
This guarantee holds under minimal assumptions and for any underlying predictive model, including complex black-box algorithms. Consequently, conformal prediction has gained traction in a wide range of real-world applications due to its flexibility, robustness, and finite-sample validity. Notable applications include survival data analysis \citep{candes2023conformalized}, causal inference \citep{yin2024conformal}, outlier detection \citep{bates2023testing}, risk control \citep{angelopoulos2024conformal}, time series forecasting \citep{angelopoulos2023conformal}, and selective inference \citep{bao2024selective}, among others.

At the same time, as the volume of data increases, the prevalence of missing data also rises. Missing data are ubiquitous in modern statistics. For example, in healthcare, discrepancies can occur when data collected from different clinics or hospitals reflect variations in the measured variables. Other contributors to missing data include sensor malfunctions, data censoring, and privacy concerns, among others. This paper focuses on missingness in covariates. In this context, we denote by $M$ the mask associated with the covariate $X$, which will be formally defined in the following section. It is crucial to ensure that the coverage guarantee holds conditional on the mask,
\begin{equation*}
	\mathbb{P}\left\{Y \in \mathcal{C}_{\alpha}(X)\mid M\right\} \geq 1 - \alpha.
\end{equation*}
From a fairness perspective, each subgroup defined by a specific mask should receive a prediction interval with equal probability of containing the true response. Moreover, missingness patterns may correlate with unobserved external features linked to meaningful categories. For example, in survey data, affluent individuals are often less likely to disclose their income, leading to missing income variables. As a result, these masks may relate to socio-economic information \citep{zaffran2023conformal}.

The approach to handling missing data depends on the specific downstream task. This topic has been extensively studied in various statistical contexts \citep{little2019statistical}.
Recent works study missing data in several different areas, including principal component analysis \citep{zhu2022high}, \textit{U}-statistics \citep{cannings2022correlation}, changepoint detection \citep{follain2022high}, testing whether the missingness is independent of the data \citep{berrett2023optimal} and classification \citep{sell2024nonparametric}. Moreover, it also receives a lots of attention in the framework of prediction \citep{le2020neumiss,le2020linear,le2021sa,ayme2022near,ayme2023naive,gui2023conformalized,shao2023distribution,seedat2023improving,zaffran2023conformal,zaffran2024predictive,josse2024consistency}.

Our work on uncertainty quantification with missing data builds builds upon the foundational work of \cite{zaffran2023conformal,zaffran2024predictive}, which examine the predictive framework of estimating \( Y \) given \( X \), where \( X \) may exhibit missing values during both training and testing phases. In \cite{zaffran2023conformal}, two conformal methods were proposed, named \textit{CQR-MDA-Exact} and \textit{CQR-MDA-Nested}, both of which achieve marginal and mask-conditional validity (defined in \eqref{marginal validity} and \eqref{mask-conditional validity} below) under specific conditions. Building on this, \cite{zaffran2024predictive} further provided a comprehensive framework in which CQR-MDA-Exact and CQR-MDA-Nested are established as special cases. 
However, the theoretical implications of missingness on marginal and mask-conditional validity remain somewhat ambiguous. Additionally, the assumption related to the missingness mechanism is a bit restrictive.  Note that \cite{sell2024nonparametric} recently proposed a general missingness mechanism specifically for binary classification tasks. Motivated by \cite{sell2024nonparametric}, we adapt this mechanism for the regression context and build upon the concept of nonexchangeable conformal prediction, as laid out by \cite{barber2023conformal}, to investigate the influence of missingness on both marginal and mask-conditional validity. 

In this article, we consider conformal prediction in the presence of missing values within the covariates. Our focus is on understanding how such missingness affects the validity and efficiency of conformal prediction. To effectively address the challenges posed by missing covariates, we introduce two novel conformal methods. The key contributions of this work are summarized as follows: (1) We show that the coverage probabilities associated with various missing data patterns differ and theoretically establish that the coverage probability is ultimately constrained by a summation of total variance. (2) We introduce a nonexchangeable conformal prediction method for missing covariates. This approach selects applicable calibration cases and assigns weights based on a distance metric that reflects similarity in missingness. We prove that the method is both marginally and mask-conditionally valid. (3) While the nonexchangeable method does not guarantee asymptotic conditional validity, we further propose a localized conformal method using kernel smoothing and a refined score function. This method is shown to be marginally valid, mask-conditionally valid, and asymptotically conditionally valid under certain assumptions.

The remainder of this article is organized as follows. Section \ref{Prelimminaries} introduces the background of conformal prediction with missing covariates and demonstrates that while conformal prediction ensures marginal validity, this alone is insufficient for comprehensive uncertainty quantification. In Section \ref{Conformal prediction with missing covariates}, we analyze the impact of missing values on conformal prediction within the framework of nonexchangeable conformal prediction. We then propose a nonexchangeable conformal prediction method for missing covariates and establish its marginal and mask-conditional validity. Later, we propose a localized conformal prediction method for missing covariates and demonstrate its marginal validity, mask-conditional validity, and asymptotical conditional validity. Section \ref{Simulation studies} presents extensive simulation studies to assess the empirical performance of the proposed methods, while Section \ref{Real data analysis} offers a real data example to illustrate their effectiveness in practice. Section \ref{Conclusion} concludes the article. All technical proofs are provided in the Appendix.

\section{Preliminaries}\label{Prelimminaries}
For convenience, some notations are also introduced. For $n\in\mathbb{N}$, let $[n]=\{1,\cdots,n\}$. Denote $|\mathcal{I}|$ as the cardinality of the index set $\mathcal{I}$. For $m,\tilde{m}\in\{0,1\}^{d}$, $m\preceq\tilde{m}$ means that $m_{i}\leq\tilde{m}_{i}, i\in [d]$, where $m_{i}$ (resp. $\tilde{m}_{i}$) is the $i$th component of the vector $m$ (resp. $\tilde{m}$). For two random variables $X$ and $Y$, $X\perp \!\!\! \perp Y$ means that $X$ and $Y$ are independent, and $X\overset{\text{d}}{=}Y$ means that $X$ and $Y$ have the same distribution.

Consider a data set with $n$ independent and identically distributed (i.i.d.) realizations of the random variable $(X,M,Y)\in\mathbb{R}^{d}\times\{0,1\}^{d}\times\mathbb{R}:(X^{(i)},M^{(i)},Y^{(i)}), i\in [n]$, where $X$ represents the covariates, $M$ the missing pattern (also called mask), and $Y$ the response to predict. For each realization $(X,M,Y)$, $M_{j}=0$ indicates that $X_{j}$ is observed, and $M_{j}=1$ indicates that $X_{j}$ is missing, i.e., \textit{NA} (Not Available). We denote by $\text{obs}(M)\subset [d]$ the indices corresponding to the observed covariates, so that $X_{\text{obs}(M)}$ corresponds to the entries actually observed. For example, if we observed $(\textit{NA},2,5)$, then $M=(1,0,0)$ and $X_{\text{obs}(M)}=(2,5)$. In the presence of missing values, the training data are $(X_{\text{obs}(M^{(i)})}^{(i)},M^{(i)},Y^{(i)})$, for $i\in[n]$, and our objective is to build a prediction set for the new response $Y^{(n+1)}$ given $X_{\text{obs}(M^{(n+1)})}^{(n+1)}$ and $M^{(n+1)}$. 

We consider missingness in the covariates, and the missingness mechanisms can be categorized into three types \citep{rubin1976inference} based on the relationship between missing pattern and observed variables: (1) \textit{Missing completely at random (MCAR)}: $M$ is independent of $X$; (2) \textit{Missing at random (MAR)}: $M$ depends only on the observed variables $X_{\text{obs}(M)}$; (3) \textit{Missing not at random (MNAR)}: neither MCAR or MAR holds. 

In the context of missing covariates, a prediction interval $\hat{\mathcal{C}}_{\alpha}$ should possesses \textit{marginal validity} as expressed by
\begin{equation}\label{marginal validity}
	\mathbb{P}\left\{Y^{(n+1)}\in \hat{C}_{\alpha}\left(X^{(n+1)},M^{(n+1)}\right)\right\}\geq 1-\alpha.
\end{equation}
Additionally, it should exhibit \textit{mask-conditional validity} such that, for any $m\in\{0,1\}^{d}$,
\begin{equation}\label{mask-conditional validity}
	\mathbb{P}\left\{Y^{(n+1)}\in \hat{C}_{\alpha}\left(X^{(n+1)},m\right)\mid M^{(n+1)}=m\right\}\geq 1-\alpha.
\end{equation}
Marginal validity (\ref{marginal validity}) indicates that the coverage guarantee is satisfied for all data across different masks as a collective group. Although marginal validity is a desirable property of a prediction interval, achieving mask-conditional validity in a distribution-free and finite-sample context is equally important. The masks create meaningful categories, and ensuring mask-conditional validity (\ref{mask-conditional validity}) promotes more equitable treatment \citep{zaffran2023conformal,zaffran2024predictive}.

Since most statistical models and machine learning methods are not inherently designed to handle incomplete data, a common approach to utilizing existing pipelines in the presence of missing values is to impute the data, thereby creating a complete dataset. We will utilize an imputation function that maps observed values to themselves while assigning plausible values to the missing entries based on those observed values.
Following the notations defined in \cite{zaffran2023conformal,zaffran2024predictive}, we define the imputation function \(\phi^{M}:\mathbb{R}^{|\text{obs}(M)|} \rightarrow \mathbb{R}^{|\text{mis}(M)|}\), where the input consists of the observed values and the output comprises the imputed values, conditional on a mask \(M \in \{0,1\}^d\). The imputation function belongs to the function class defined as follows:
\[
\mathcal{F}^{I} := \left\{ \Phi: \mathbb{R}^{d} \times \{0,1\}^{d} \to \mathbb{R}^{d} : \Phi_{i}(X,M) = X_{i} \text{1}_{M_{i}=0} + \phi_{i}^{M}(X_{\text{obs}(M)}) \text{1}_{M_{i}=1}, \, i \in [d] \right\}.
\]
In this formulation, \(\Phi_{i}(X,M)\) assigns the original observed value \(X_{i}\) for indices where \(M_{i} = 0\) (indicating the presence of an observed value), while for indices where \(M_{i} = 1\) (indicating a missing value), it applies the imputation function \(\phi_{i}^{M}(X_{\text{obs}(M)})\) to generate plausible values based on the observed data. This approach facilitates the integration of imputed datasets into standard analytical frameworks.

Suppose we have an additional independent training dataset drawn from the same distribution, denoted as \(\{(X^{(i)}, M^{(i)}, Y^{(i)}), i \in \mathcal{I}_{\text{train}}\}\), which can be used to fit both the imputation function and the regression function that we will discuss subsequently. This approach is known as \textit{inductive conformal prediction} \citep{vovk2005algorithmic} or \textit{split conformal prediction} \citep{lei2018distribution}. From the perspective of split conformal prediction, there are \( |\mathcal{I}_{\text{train}}| \) data points utilized during the training phase and \( n \) data points in the calibration phase. In practice, if only \( n \) data points are available in total, we might allocate half of them, \( n/2 \), to create the training dataset, reserving the remaining data for calibration. For clarity and consistency throughout the paper, we will continue to refer to the calibration set size as \( n \) for the split conformal method, facilitating universal notation across different methodologies.

Similar to the approach described in \cite{barber2023conformal}, the imputation function can be learned through an imputation algorithm represented as follows:
\begin{equation}\label{imputation algorithm}
	\mathcal{A}_{1}: \left(\mathbb{R}^{d}\times\{0,1\}^{d}\right)^{|\mathcal{I}_{train}|}\rightarrow\left\{\text{imputation function }\hat{\Phi}:\mathbb{R}^{d}\times\{0,1\}^{d}\rightarrow\mathbb{R}^{d}\right\},
\end{equation}
which maps a dataset containing any number of pairs \((X, M)\) to a fitted imputation function \(\hat{\Phi}\). Similarly, the regression function can be learned from a regression algorithm defined as:
\begin{equation}\label{regression algorithm}
	\mathcal{A}_{2}: \left(\mathbb{R}^{d}\times\mathbb{R}\right)^{|\mathcal{I}_{\text{train}}|}\rightarrow\left\{\text{regression function }\hat{\mu}: \mathbb{R}^{d}\rightarrow\mathbb{R}\right\},
\end{equation}
which maps a complete imputed dataset to a fitted regression function \(\hat{\mu}\). Consequently, the response variable \(Y\) can be estimated or predicted by first imputing the data to create a complete dataset and then applying the regression function to the imputed data, expressed as:
\begin{equation*}
	\hat{Y}=\hat{\mu}\circ\hat{\Phi}(X,M),
\end{equation*}
where \(\hat{\mu} \circ \hat{\Phi}\) represents a composite function that effectively combines both imputation and prediction processes.

We then construct a prediction interval for the response $Y$. Define the score function 
\begin{equation}\label{score function}
	V:=V(X,M,Y)=|Y-\hat{\mu}\circ\hat{\Phi}(X,M)|,
\end{equation}
and then compute the scores of the $n$ data points, i.e., 
\begin{equation*}
	V^{(i)}=|Y^{(i)}-\hat{\mu}\circ\hat{\Phi}(X^{(i)},M^{(i)})|,\quad i\in [n].
\end{equation*}
Denote $Q_{1-\alpha}(F)$ as the level $(1-\alpha)$ quantile of a distribution $F$, i.e., for $Z\sim F$,
\begin{equation*}
	Q_{1-\alpha}(F)=\inf\left\{z:\mathbb{P}\{Z\leq z\}\geq 1-\alpha\right\}.
\end{equation*}
The prediction interval at the new covariate and missing indicator $(X^{(n+1)},M^{(n+1)})$ can be written as
\begin{equation}\label{naive conformal}
	\hat{\mathcal{C}}_{n}(X^{(n+1)},M^{(n+1)})=\left\{y\in\mathbb{R}: V(X^{(n+1)},M^{(n+1)},y)\leq Q_{1-\alpha}\left(\sum_{i=1}^{n}\frac{1}{n+1}\cdot\delta_{V^{(i)}}+\frac{1}{n+1}\cdot\delta_{+\infty}\right)\right\},
\end{equation}
where $\delta_{a}$ is a point mass at $a$ (i.e., the distribution that places all mass at the value $a$). 

\begin{assumption}[Symmetrical imputation and regression algorithms]\label{symmetry}
	The imputation algorithm $\mathcal{A}_{1}$ in (\ref{imputation algorithm}) and regression algorithm $\mathcal{A}_{2}$ in (\ref{regression algorithm}) treat their input data points symmetrically, i.e., 
	\begin{equation*}
		\mathcal{A}_{1}\left((X^{(i)},M^{(i)}),i\in [n]\right)=\mathcal{A}_{1}\left((X^{(\sigma(i))},M^{(\sigma(i))}),i\in [n]\right),
	\end{equation*}
	and
	\begin{equation*}
		\mathcal{A}_{2}\left((\hat{\Phi}(X^{(i)},M^{(i)}),Y^{(i)}), i\in [n]\right)=\mathcal{A}_{2}\left((\hat{\Phi}(X^{(\sigma(i))},M^{(\sigma(i))}),Y^{(i)}), i\in [n]\right),
	\end{equation*}
	for all $n\geq 1$, and all permutation $\sigma$ on $[n]$, where $\hat{\Phi}$ is the imputation function learned by the imputation algorithm.
\end{assumption}

Symmetrical condition is a mild condition relative to the regression algorithm \citep{vovk2005algorithmic,lei2018distribution,barber2023conformal}. In addition, most existing imputation methods for exchangeable (a condition weaker than i.i.d.) data satisfy the symmetrical condition \citep{zaffran2023conformal,zaffran2024predictive}. 

\begin{theorem}[Marginal validity \citep{zaffran2023conformal,zaffran2024predictive}]\label{Marginal validity}
	Given i.i.d. observations $(X^{(i)},M^{(i)},Y^{(i)}), i\in [n]$, and suppose that Assumption \ref{symmetry} holds. For a new i.i.d. observation $(X^{(n+1)},M^{(n+1})$, the conformal prediction interval $\hat{\mathcal{C}}_{n}$ defined in (\ref{naive conformal}) is marginally valid, i.e., 
	\begin{equation*}
		\mathbb{P}\left\{Y^{(n+1)}\in\hat{\mathcal{C}}_{n}\left(X^{(n+1)},M^{(n+1)}\right)\right\}\geq 1-\alpha.
	\end{equation*}
\end{theorem}

By the i.i.d. and symmetrical algorithms assumptions, the scores are exchangeable. This result shows that marginal validity holds regardless of the missing mechanism (MCAR, MAR or MNAR) and the symmetric imputation scheme. The prediction interval (\ref{naive conformal}) is marginally valid (\ref{marginal validity}), under mild conditions; however, achieving mask-conditional validity (\ref{mask-conditional validity}) is a challenging task. To better illustrate the interplay between missing values and uncertainty quantification with respect to the mask, we consider the following example.

\begin{example}\label{example}
	Consider the following data-generating process:
	\begin{equation*}
		Y = \beta_1 X_1 + \beta_2 X_2 + \varepsilon,
	\end{equation*}
	where \( \varepsilon \sim N(0, \sigma^2) \) is independent of both \( X \) and \( M \), \( \beta_1, \beta_2 \) are scalars, and
	\begin{equation*}
		\begin{pmatrix}
			X_1 \\ X_2
		\end{pmatrix}
		\sim N \left(
		\begin{pmatrix}
			\mu_1 \\ \mu_2
		\end{pmatrix},
		\begin{pmatrix}
			\sigma_1^2 & \rho \sigma_1 \sigma_2 \\
			\rho \sigma_1 \sigma_2 & \sigma_2^2
		\end{pmatrix}
		\right).
	\end{equation*}
	
	The missing data mechanism is defined as follows: \( M_1 = 1 \) if \( X_1 > \tau_1 \), and \( M_2 = 1 \) if \( X_2 < \tau_2 \). Then, the conditional variance of the response variable given the observed covariates and a specific mask \( m \) is
	\begin{equation*}
		\text{Var}(Y \mid X_{\text{obs}(m)}, M = m) = 
		\begin{cases}
			\sigma^2, & m = (0,0) \\[6pt]
			\beta_1^2 \sigma_{1|2}^2 \left(
			1 + \dfrac{\frac{\tau_1 - \mu_{1|2}}{\sigma_{1|2}} \, \phi\left( \frac{\tau_1 - \mu_{1|2}}{\sigma_{1|2}} \right)}{1 - \Phi\left( \frac{\tau_1 - \mu_{1|2}}{\sigma_{1|2}} \right)}
			- \left[ \dfrac{ \phi\left( \frac{\tau_1 - \mu_{1|2}}{\sigma_{1|2}} \right) }{1 - \Phi\left( \frac{\tau_1 - \mu_{1|2}}{\sigma_{1|2}} \right)} \right]^2
			\right) + \sigma^2, & m = (1,0) \\[10pt]
			\beta_2^2 \sigma_{2|1}^2 \left(
			1 - \dfrac{ \frac{\tau_2 - \mu_{2|1}}{\sigma_{2|1}} \, \phi\left( \frac{\tau_2 - \mu_{2|1}}{\sigma_{2|1}} \right)}{ \Phi\left( \frac{\tau_2 - \mu_{2|1}}{\sigma_{2|1}} \right)}
			- \left[ \dfrac{ \phi\left( \frac{\tau_2 - \mu_{2|1}}{\sigma_{2|1}} \right) }{ \Phi\left( \frac{\tau_2 - \mu_{2|1}}{\sigma_{2|1}} \right)} \right]^2
			\right) + \sigma^2, & m = (0,1)
		\end{cases}
	\end{equation*}
	where
	\[
	\mu_{1|2} = \mu_1 + \rho \frac{\sigma_{1}}{\sigma_{2}}(X_2 - \mu_2), \quad
	\sigma_{1|2}^2 = \sigma_1^2 (1 - \rho^2), \quad
	\mu_{2|1} = \mu_2 + \rho \frac{\sigma_2}{\sigma_{1}} (X_1 - \mu_1), \quad
	\sigma_{2|1}^2 = \sigma_2^2 (1 - \rho^2),
	\]
	and \( \phi(\cdot) \) and \( \Phi(\cdot) \) denote the probability density function and cumulative distribution function of the standard normal distribution, respectively.
\end{example}

Example \ref{example} emphasizes that even when the noise of the generative model is homoscedastic, missing values induce heteroskedasticity. Indeed, the variance of the conditional distribution of $Y\mid(X_{\text{obs}(m)},M=m)$ depends on missing pattern $m$. 

\section{Conformal prediction with missing covariates}\label{Conformal prediction with missing covariates}
In this section, we propose two methods of conformal prediction with missing covariates.
\subsection {Nonexchangeable conformal prediction}
The prediction interval in (\ref{naive conformal}) can be written equivalently as 
\begin{equation*}
	\hat{\mathcal{C}}_{n}(X^{(n+1)},M^{(n+1)})=\hat{\mu}\circ\hat{\Phi}(X^{(n+1)},M^{(n+1)})\pm\left(\text{the $\lceil (1-\alpha)(n+1)\rceil$-th smallest of $V^{(i)},i\in [n]$}\right),
\end{equation*}
which means that the interval length remains invariant regardless of the mask. However, Example \ref{example} shows that the prediction interval length with respect to different masks should be variant. As missing values induce heteroskedasticity, to build a prediction interval for $Y^{(n+1)}$ given $X_{\text{obs}(m)}^{(n+1)}$ and $M^{(n+1)}=m$, a natural choice is to upweight the data with missingness similar to the test point and downweight those with missingness not similar to the test point. 

Define $d^{(i)}:=d(X^{(i)},X^{(n+1)})$ as the distance metric between $X^{(i)}$ and $X^{(n+1)}$. For example, the distance can be Heterogeneous Euclidean-Overlap Metric (HEOM), Heterogeneous Value Difference Metric (HVDM) \citep{wilson1997improved}, etc. Then, the scores and distances can be calculated, $(V^{(i)},d^{(i)}),i\in [n]$. Define $w_{i}\in [0,1],i\in [n]$, a large $d^{(i)}$ corresponds to a small $w_{i}$. Suppose that $d^{(i)},i\in [n]$ are arranged in ascending order, i.e., $d^{(1)}\geq\cdots\geq d^{(n)}$, and hence, $w_{1}\leq\cdots\leq w_{n}$. Following the argument in \cite{barber2023conformal}, we can set $w_{i}=\rho^{n+1-i},i\in [n+1]$, e.g., $\rho=0.99$. The nonexchangeable conformal prediction interval is defined as  
\begin{equation}\label{nonexchangeable conformal prediction}
	\hat{\mathcal{C}}_{n}(X^{(n+1)},M^{(n+1)})=\left\{y\in\mathbb{R}:V^{(n+1)}\leq Q_{1-\alpha}\left(\sum_{i=1}^{n}\tilde{w}_{i}\delta_{V^{(i)}}+\tilde{w}_{n+1}\delta_{+\infty}\right)\right\},
\end{equation}
where $\tilde{w}_{i}=w_{i}/(w_{1}+\cdots+w_{n+1}),i\in [n+1]$.
Let the residual vector $R\in \mathbb{R}^{n+1}$ with entries 
\begin{equation*}
	R_{i}:=\left|Y^{(i)}-\hat{\mu}\circ\hat{\Phi}(X^{(i)},M^{(i)})\right|=V^{(i)},\quad i\in [n+1].
\end{equation*}
Denote $R^{i}$ as this sequence after swapping the test score $V^{(n+1)}$ with the $i$-th training score, i.e.,
\begin{equation*}
	R^{i}=\left(V^{(1)},\cdots,V^{(i-1)},V^{(n+1)},V^{(i+1)},\cdots,V^{(n)},V^{(i)}\right).
\end{equation*}

\begin{lemma}\label{lemma:nonexchangeable conformal prediction}
	The nonexchangeable conformal prediction interval defined in (\ref{nonexchangeable conformal prediction}) satisfies 
	\begin{equation*}
		\mathbb{P}\left\{Y^{(n+1)}\in\hat{\mathcal{C}}_{n}\left(X^{(n+1)},M^{(n+1)}\right)\right\}\geq 1-\alpha-\sum_{i=1}^{n}\tilde{w}_{i}\cdot\text{d}_{\text{TV}}\left(R,R^{i}\right),
	\end{equation*}
	where $\text{d}_{\text{TV}}(\cdot,\cdot)$ is the total variance between two distributions.
\end{lemma}

\begin{remark}
	The similar result of the nonexchangeable conformal prediction without considering missingness can be found in \cite{barber2023conformal}. Note that Lemma \ref{lemma:nonexchangeable conformal prediction} does not require the exchangeability of $\left(X^{(i)},M^{(i)},Y^{(i)}\right),i\in [n+1]$. The weights $w_{i}$ in the nonexchangeable conformal method are fixed, and we only need qualitative knowledge of the likely deviation from exchangeability. For example, we can use HEOM \citep{wilson1997improved} to determine the weights. HEOM is widely used as $k$-nearest neighbors imputation, which has good predictive performance, see e.g., \cite{santos2020distance}. With a good distance metric, the distance between two data points with similar missing patterns would be smaller than those without similar missing patterns. Thus, our result above confirms the intuition that we should give higher weights $w_{i}$ to data points $\left(X^{(i)},M^{(i)},Y^{(i)}\right)$ that we believe are drawn from a similar distribution as $\left(X^{(n+1)},M^{(n+1)},Y^{(n+1)}\right)$, and lower weights to those that are less reliable.
\end{remark}

By Lemma \ref{lemma:nonexchangeable conformal prediction}, the following Theorem is straightforward.

\begin{theorem}\label{theorem:nonexchangeable conformal prediction}
	Given i.i.d. observations $(X^{(i)},M^{(i)},Y^{(i)}), i\in [n]$, and suppose that Assumption \ref{symmetry} holds.  For a new i.i.d. observation $(X^{(n+)},M^{(n+1)})$, the nonexchangeable conformal prediction interval $\hat{\mathcal{C}}_{n}$ defined in (\ref{nonexchangeable conformal prediction}) is marginally valid, i.e.,
	\begin{equation*}
		\mathbb{P}\left\{Y^{(n+1)}\in\hat{\mathcal{C}}_{n}\left(X^{(n+1)},M^{(n+1)}\right)\right\}\geq 1-\alpha.
	\end{equation*}
	Meanwhile, 
	\begin{equation*}
		\mathbb{P}\left\{Y^{(n+1)}\in\hat{\mathcal{C}}_{n}\left(X^{(n+1)},m\right)\mid M^{(n+1)}=m\right\}\geq 1-\alpha-\mathop{\sum}_{i\in [n]:M^{(i)}\neq m}\tilde{w}_{i}\cdot\text{d}_{\text{TV}}\left(R,R^{i}\right).
	\end{equation*}
\end{theorem}

\begin{remark}
	In the context of the marginal validity of the theorem, the data points \((X^{(i)},M^{(i)},Y^{(i)})\) for \(i \in [n+1]\) are indeed independent and identically distributed (i.i.d.). Consequently, the associated scores \(V^{(i)}\) for \(i \in [n+1]\) exhibit exchangeability. This leads to the equivalence in distribution \(R \overset{\text{d}}{=} R^{i}\), and it follows that the total variation distance \(\text{d}_{\text{TV}}(R, R^{i}) = 0\) holds for all \(i\). Importantly, the introduction of fixed weights \(w_{i}\) into the quantile calculations for constructing the prediction interval does not result in any loss of coverage with respect to marginal validity.
	
	Conversely, with respect to the mask-conditional probability addressed in the theorem, the condition of probability being dependent on the mask implies that the data points may lack the property of exchangeability. Specifically, while these points may be independent, they are not necessarily identically distributed. Although the overall data points originate from the same distribution, distinct subgroups defined by different masks may have disparate distributions. For example, as delineated in Example \ref{example}, data corresponding to different masks may indeed arise from different distributions.
	Moreover, violations of exchangeability manifest due to the conditional variance of \(Y \mid (X_{\text{obs}(M)}, M)\) varying with different masks \(M\). As a result, the total variation distance \(\text{d}_{\text{TV}}(R, R^{i}) = 0\) holds for data points indexed by \(i\) that share the same mask as the test point, whereas this distance may not be zero for data points associated with different masks.
	
\end{remark}

Beginning with nonexchangeable conformal prediction, we demonstrate that achieving marginal validity is feasible. However, attaining mask-conditional validity presents considerable challenges. As outlined in Theorem \ref{theorem:nonexchangeable conformal prediction}, a simplistic approach to achieve mask-conditional validity would be to exclude data points indexed by \(i\) such that \(M^{(i)} \neq M^{(n+1)}\), retaining only those data points that share the same mask as the test point. This strategy, however, proves unrealistic in scenarios characterized by missing covariates.
To illustrate this, consider the missing pattern \(m \in \{0,1\}^{d}\). The number of distinct missing patterns tends to grow exponentially with the dimensionality \(d\); in fact, there are \(2^{d}-1\) possible missing patterns. Consequently, the quantity of data available for any specific missing pattern may be exceedingly limited \citep{zaffran2023conformal,zaffran2024predictive}.
When reassessing the conditional probability, a more reasonable strategy for achieving mask-conditional validity is to retain some data points where the total variance can be made zero under certain mild conditions, while discarding others that cannot satisfy this requirement. To support this approach, we will introduce some general assumptions about the missingness mechanism.

\begin{assumption}\label{general assumption}
	For all $X\in\mathbb{R}^{d},m,\tilde{m}\in\{0,1\}^{d}$ with $\tilde{m}\preceq m$,  
	\begin{equation}\label{general assumption_1}
		Y\mid\left(X_{\text{obs}(m)}\right)\overset{\text{d}}{=}Y\mid\left(X_{\text{obs}(m)},M=\tilde{m}\right).
	\end{equation}
	Equivalently,
	\begin{equation*}
		Y\mid\left(X_{\text{obs}(m)},M=m\right)\overset{\text{d}}{=}Y\mid\left(X_{\text{obs}(m)},M=\tilde{m}\right).
	\end{equation*}
	
\end{assumption}

\begin{remark}
	To understand the generality afforded by Assumption \ref{general assumption}, first note that (\ref{general assumption_1}) holds trivially if $M$ is independent of $(X,Y)$. Besides, if $Y\perp \!\!\! \perp M\mid X_{\text{obs}(m)}$, then Assumption \ref{general assumption} is also satisfied. This general assumption on the missingness mechanism is first proposed in \cite{sell2024nonparametric}, where they consider classification problems, i.e., $Y\in \{0,1\}$. They point out that this assumption includes where $M$ and $(X,Y)$ are dependent, and may even allow the data to be MNAR.  
\end{remark}

We say that $(X,M,Y)$ is an \textit{available case} for $m$, if $M\preceq m$. Then the available cases for the mask of the test point $\left(X^{(n+1)},M^{(n+1)},Y^{(n+1)}\right)$ can be indexed by $\tilde{\mathcal{I}}=\{i\in [n]:M^{(i)}\preceq M^{(n+1)}\}$. Under Assumption \ref{general assumption}, the conditional distribution of $Y\mid\left(X_{\text{obs}(M^{(n+1)})},M^{(i)}\right)$, for $i\in\tilde{\mathcal{I}}$, will be equal to that of $Y\mid\left(X_{\text{obs}(M^{(n+1)})},M^{(n+1)}\right)$. Algorithm \ref{algorithm:nexCP_new} presents a new approach to nonexchangeable conformal prediction.

\begin{algorithm}[ht]
	\caption{Nonexchangeable conformal prediction with missing covariates}
	\begin{algorithmic}[1]
		\State\textbf{Input}: Imputation algorithm $\mathcal{A}_{1}$, regression algorithm $\mathcal{A}_{2}$, distance function $d(\cdot,\cdot)$, miscoverage level $\alpha\in(0,1)$, $\rho\in(0,1)$, training data $\left\{\left(X^{(i)},M^{(i)},Y^{(i)}\right),i\in\mathcal{I}_{\text{train}}\right\}$, calibration data $\left\{\left(X^{(i)},M^{(i)},Y^{(i)}\right),i\in[n]\right\}$, new observation $\left(X^{(n+1)},M^{(n+1)}\right)$.
		\State\textbf{Process}
		\State Learn the imputation function: $\hat{\Phi}(\cdot,\cdot)\gets\mathcal{A}_{1}\left(\left\{\left(X^{(i)},M^{(i)}\right),i\in\mathcal{I}_{\text{train}}\right\}\right)$
		\State Learn the regression function: $\hat{\mu}(\cdot)\gets\mathcal{A}_{2}\left(\left\{\left(\hat{\Phi}\left(X^{(i)},M^{(i)}\right),Y^{(i)}\right),i\in\mathcal{I}_{\text{train}}\right\}\right)$
		\State Find available cases $\tilde{\mathcal{I}}=\left\{i\in[n]:M^{(i)}\preceq M^{(n+1)}\right\}$
		\State Compute the scores as $\tilde{V}^{(i)}=\left|Y^{(i)}-\hat{\mu}\circ\hat{\Phi}\left(X^{(i)},M^{(n+1)}\right)\right|,i\in\tilde{\mathcal{I}}$
		\State Compute the distances as $d^{(i)}=d\left(X^{(i)},X^{(n+1)}\right),i\in\tilde{\mathcal{I}}$
		\State Rearrange the scores based on the ascending order of the distances $d^{(i)},i\in\tilde{\mathcal{I}}$. The scores are denoted as $\tilde{V}_{*}^{(i)},i\in[|\tilde{\mathcal{I}}|]$
		\State Calculate $w_{i}=\begin{cases}
			\rho^{\left|\tilde{\mathcal{I}}\right|+1-i},& i \in [|\tilde{\mathcal{I}}|]\text{ and } M^{(i)}\neq M^{(n+1)}, \\
			1, & i \in [|\tilde{\mathcal{I}}|]\text{ and } M^{(i)}=M^{(n+1)}. 
		\end{cases}$, and $w_{n+1}=1$. Normalize the weights: $\tilde{w}_{i}=w_{i}/\left(w_{n+1}+{\textstyle \sum_{j\in[|\tilde{\mathcal{I}}|]}}w_{j}\right),i\in[|\tilde{\mathcal{I}}|]$, and $\tilde{w}_{n+1}=w_{n+1}/\left(w_{n+1}+{\textstyle \sum_{j\in[|\tilde{\mathcal{I}}|]}}w_{j}\right)$.
		\State\textbf{Output}: let $d=Q_{1-\alpha}\left({\textstyle \sum_{i\in [|\tilde{\mathcal{I}}|]}}\tilde{w}_{i}\cdot\delta_{\tilde{V}_{*}^{(i)}}+\tilde{w}_{n+1}\cdot\delta_{+\infty}\right)$,
		\begin{equation*}
			\hat{\mathcal{C}}_{n}\left(X^{(n+1)},M^{(n+1)}\right)=\left[\hat{\mu}\circ\hat{\Phi}\left(X^{(n+1)},M^{(n+1)}\right)-d,\hat{\mu}\circ\hat{\Phi}\left(X^{(n+1)},M^{(n+1)}\right)+d\right].
		\end{equation*}
		
	\end{algorithmic}\label{algorithm:nexCP_new}
\end{algorithm}

\begin{remark}
	In Algorithm \ref{algorithm:nexCP_new}, Lines 3-4 focus on learning the imputation and regression functions from the training set. In Line 5, the algorithm identifies the available cases in the calibration set relevant to the test point. Subsequent to this, Line 6 computes the scores of the available cases, which are connected to the mask of the test point \(M^{(n+1)}\) via the formula \(\tilde{V}^{(i)} = \left|Y^{(i)} - \hat{\mu} \circ \hat{\Phi}\left(X^{(i)}, M^{(n+1)}\right)\right|\) for \(i \in \tilde{\mathcal{I}}\). For instance, if \(M^{(n+1)} = (1, 0, 0)\), and \(X^{(i)}\) is fully observed (e.g., \(X^{(i)} = (-1, 2, 5)\)), then \(\hat{\Phi}\left(X^{(i)}, M^{(n+1)}\right)\) would account for the first entry of \(X^{(i)}\) as missing and impute a plausible value accordingly. Lines 7-8 are responsible for calculating the distances between the available cases and the test point to assess their similarity, and they reorder the scores based on these distances. Line 9 calculates fixed weights \(\tilde{w}_{i} \in [0, 1]\) in nonexchangeable conformal prediction, which are not data-dependent. This weight scheme gives higher weights to data points considered similar to the test point in distribution. Notably, points sharing the same mask as the test point receive the highest weights. The final prediction interval is determined by taking the \(1 - \alpha\) quantile of the empirical distribution described by \({\textstyle \sum_{i \in [|\tilde{\mathcal{I}}|]}}\tilde{w}_{i} \cdot \delta_{\tilde{V}_{*}^{(i)}} + \tilde{w}_{n+1} \cdot \delta_{+\infty}\).
	
\end{remark}

\begin{theorem}\label{theorem:nexCP_new}
	Given i.i.d. observations $(X^{(i)},M^{(i)},Y^{(i)}), i\in [n]$, and suppose Assumptions \ref{symmetry}-\ref{general assumption} hold, and the missing data mechanism is MCAR. For a new observation $(X^{(n+)},M^{(n+1)})$, the nonexchangeable conformal prediction interval defined in Algorithm \ref{algorithm:nexCP_new} is marginally valid,
	\begin{equation*}
		\mathbb{P}\left\{Y^{(n+1)}\in\hat{\mathcal{C}}_{n}\left(X^{(n+1)},M^{(n+1)}\right)\right\}\geq 1-\alpha.
	\end{equation*}
	Moreover, it is also mask-conditionally valid, i.e., for $m\in\{0,1\}^{d}$,
	\begin{equation*}
		\mathbb{P}\left\{Y^{(n+1)}\in\hat{\mathcal{C}}_{n}\left(X^{(n+1)},m\right)\mid M^{(n+1)}=m\right\}\geq 1-\alpha.
	\end{equation*}
\end{theorem}

\begin{remark}
	
	Under Assumption \ref{general assumption}, we demonstrate the mask-conditional validity of the proposed nonexchangeable conformal prediction. In Line 6 of Algorithm \ref{algorithm:nexCP_new}, \(\hat{\Phi}(X^{(i)}, M^{(n+1)})\) imputes plausible values based on the mask of the test point \(M^{(n+1)}\). This is akin to the \textit{Missing Data Augmentation} (MDA) strategy discussed in \cite{zaffran2023conformal}.
	In practice, if we set all the weights to be equal and use the conformalized quantile regression score \citep{romano2019conformalized}, our approach to the nonexchangeable conformal prediction interval would yield results similar to the \textit{CQR-MDA-Exact} method from \cite{zaffran2023conformal}. However, our starting point is different: we begin with nonexchangeable conformal prediction.
	We also use a mild assumption about the missingness mechanism (Assumption \ref{general assumption}), unlike the stronger condition implicitly required for CP-MDA-Exact in \cite{zaffran2023conformal}, such as \(M \perp \!\!\! \perp (X,Y)\), which is just a particular case within our general framework for missingness.
	Additionally, the distance metric provides valuable information, and the weights play an important role in our nonexchangeable conformal prediction. This could improve the efficiency of constructing the prediction interval.
	
\end{remark}

In practice, it is also important to achieve the strongest coverage guarantee conditional on both $X$ and $M$, i.e., for $x\in\mathbb{R}^{d}$ and $m\in\{0,1\}^{d}$,
\begin{equation}\label{conditonal_on_X}
	\mathbb{P}\{Y^{(n+1)}\in\hat{C}_{\alpha}(x,m)\mid X^{(n+1)}=x,M^{(n+1)}=m\}\geq 1-\alpha.
\end{equation}
While it is impossible to achieve informative prediction interval satisfying coverage guarantee conditional on the covariate $X$ in distribution-free settings, such conditional coverage can be achieved asymptotically, under suitable regularity assumptions \citep{lei2014distribution,barber2021limits}. In contrast, Theorem \ref{theorem:nexCP_new} states that the proposed nonexchangeable conformal prediction is marginally valid and mask-conditionally valid but not asymptotical conditional valid. In the following, we introduce a conformal method that satisfies all three properties: marginal validity, mask-conditional validity, and asymptotic conditional validity.

\subsection{Efficient conformal prediction by localization}

In the construction of nonexchangeable conformal prediction, the weights are determined by the ordering of the distance that measures the similarity of the missing covariates. Notably, nonexchangeable conformal prediction requires only qualitative knowledge regarding the distance of the missing covariates. This approach ensures that data points with similar missing patterns receive higher weights. 
However, if we leverage the quantitative information provided by the distance, we may enhance the prediction efficiency by utilizing this additional information. 

To achieve improved prediction efficiency, we propose localized conformal prediction, which harnesses the quantitative insights derived from the distance. Localized conformal prediction is adept at addressing issues such as data heteroscedasticity and potential covariate shift. Recent studies in this area include works by \cite{han2022split, guan2023localized, mao2024valid, hore2024conformal}. 
The core idea behind localized conformal prediction is to define a localizer surrounding the covariate of the test point, thereby up-weighting samples that are in proximity to the test point, based on the localizer. This method leads to more homogeneous quantiles across various covariates in the weighted distribution of the score. 
In the following sections, we will introduce the concept of localized conformal prediction specifically in the context of missing covariates.

Recall the score function in (\ref{score function}), i.e.,
\begin{equation*}
	V:=V(X,M,Y)=|Y-\hat{\mu}\circ\hat{\Phi}(X,M)|.
\end{equation*}
If we condition $X=X^{(n+1)}.M=M^{(n+1)}$, then $V$ can be considered as a random variable of $Y$ conditional on $X^{(n+1)}$ and $M^{(n+1)}$. Let $F\left(v\mid X=x,M=m\right)=\mathbb{P}\left\{V\leq v\mid X=x,M=m\right\}$ be the conditional cumulative distribution function (CDF) of $V$ given $X=x,M=m$. Then, a prediction interval based on the conditional CDF can be obtained as
\begin{equation*}
	\hat{\mathcal{C}}_{\alpha}\left(X^{(n+1)},M^{(n+1)}\right)=\left\{y\in\mathbb{R}:V\left(X^{(n+1)},M^{(n+1)},y\right)\leq Q_{1-\alpha}\left(F\left(v\mid X=X^{(n+1)},M=M^{(n+1)}\right)\right)\right\},
\end{equation*}
which is an aracle prediction interval with the smallest interval length and satisfying conditional validity \citep{lei2018distribution}. If we knew the real conditional distribution of $V\mid (X=x,M=m)$, the prediction interval constructed above automatically satisfies the conditional coverage probability in (\ref{conditonal_on_X}). Nonetheless, in practice, the real conditional distribution of $V\mid (X=x,M=m)$ cannot be obtained. A natural way is to plug in the real conditional distribution by an approximate estimated conditional distribution $\hat{F}\left(v\mid X=x,M=m\right)$. However, since $\hat{F}$ is just an approximation of the ground truth, this plug-in strategy generally has no coverage guarantee. Moreover, it becomes more challenging in the present of missing covariates, as we need to consider different kind of validity, i.e., marginal validity and mask-conditional validity.

We now provide a new localized conformal prediction with missing covariates. Note that Assumption \ref{general assumption} is also crucial in the localized conformal prediction. We begin with finding the available cases for $M^{(n+1)}$ in the training and calibration set, i.e., $\tilde{\mathcal{I}}_{\text{train}}=\{i\in\mathcal{I}_{\text{train}}:M^{(i)}\preceq M^{(n+1)}\}$ and $\tilde{\mathcal{I}}=\{i\in [n]:M^{(i)}\preceq M^{(n+1)}\}$. Then, the scores can be written as 
\begin{equation*}
	\tilde{V}^{(i)}:=V(X^{(i)},M^{(n+1)},Y^{(i)})=|Y^{(i)}-\hat{\mu}\circ\hat{\Phi}(X^{(i)},M^{(n+1)})|,\quad i\in\tilde{\mathcal{I}}_{\text{train}}\cup\tilde{\mathcal{I}}.
\end{equation*}
Under Assumption \ref{general assumption}, we can approximate the condition CDF of $F(v\mid X=x,M=M^{(n+!)})$ by localization with kernel smoothing as 
\begin{equation*}
	\hat{F}_{h}\left(V=v\mid X=x,M=M^{(n+1)}\right)=\sum_{i\in\tilde{\mathcal{I}}_{\text{train}}}p_{h}(X^{(i)}\mid x)\text{1}(v\leq \tilde{V}^{(i)}),
\end{equation*}
where 
\begin{equation*}
	p_{h}(X^{(i)}\mid x)=\frac{K\left(d(X^{(i)},x)/h\right)}{\sum_{j\in\tilde{\mathcal{I}}_{\text{train}}}K\left(d(X^{(j)},x)/h\right)}.
\end{equation*}
Here, $K(\cdot)$ is any kernel function, $d(\cdot,\cdot)$ is any distance metric measuring similarity of two missing covariates, $h$ is the bandwidth in kernel smoothing, and $\text{1}(\cdot)$ is the indicator function. Note that the above conditional distribution is conditional on two variables $X$ and $M$, we fix $M=M^{(n+1)}$ specifically so that only $X$ changes. We then construct $\hat{\mathcal{C}}_{n}\left(X^{(n+1)},M^{(n+1)}\right)$ based on the weighted empirical distribution $\hat{F}_{h}(V=v\mid X=x,M=M^{(n+1)})$. In particular, let
\begin{equation*}
	\hat{F}_{h}^{(i)}=\hat{F}_{h}\left(V=v\mid X=X^{(i)},M=M^{(n+1)}\right),\quad i\in\tilde{\mathcal{I}}\cup\{n+1\}.
\end{equation*}
and
\begin{equation}\label{localized score function}
	V_{\alpha,h}^{(i)}:=V_{\alpha,h}\left(X^{(i)},M^{(n+1)},Y^{(i)}\right)=\tilde{V}^{(i)}-Q_{1-\alpha}\left(\hat{F}_{h}^{(i)}\right),\quad i\in\tilde{\mathcal{I}}\cup\{n+1\}.
\end{equation}
The new localized conformal prediction interval can be constructed as
\begin{equation*}
	\hat{\mathcal{C}}_{n}\left(X^{(n+1)},M^{(n+1)}\right)=\left\{y\in\mathbb{R}:V_{\alpha,h}\left(X^{(n+1)},M^{(n+1)},y\right)\leq Q_{1-\alpha}\left(\frac{1}{|\tilde{\mathcal{I}}|+1}\sum_{i\in\tilde{\mathcal{I}}}\delta_{V_{\alpha,h}^{(i)}}+\frac{1}{|\tilde{\mathcal{I}}|+1}\delta_{+\infty}\right)\right\}.
\end{equation*}
The localized conformal prediction is thus summarized in Algorithm \ref{algorithm:lcp}. 

\begin{algorithm}[ht]
	\caption{Localized conformal prediction with missing covariates}
	\begin{algorithmic}[1] 
		\State\textbf{Input}: Imputation algorithm $\mathcal{A}_{1}$, regression algorithm $\mathcal{A}_{2}$, kernel function $K(\cdot)$, distance function $d(\cdot,\cdot)$, bandwidth $h$, miscoverage level $\alpha\in(0,1)$, training data $\left\{\left(X^{(i)},M^{(i)},Y^{(i)}\right),i\in\mathcal{I}_{\text{train}}\right\}$, calibration data $\left\{\left(X^{(i)},M^{(i)},Y^{(i)}\right),i\in[n]\right\}$, new observation $\left(X^{(n+1)},M^{(n+1)}\right)$.
		\State \textbf{Process}
		\State Learn the imputation function: $\hat{\Phi}(\cdot,\cdot)\gets\mathcal{A}_{1}\left(\left\{\left(X^{(i)},M^{(i)}\right),i\in\mathcal{I}_{\text{train}}\right\}\right)$
		\State Learn the regression function: $\hat{\mu}(\cdot)\gets\mathcal{A}_{2}\left(\left\{\left(\hat{\Phi}\left(X^{(i)},M^{(i)}\right),Y^{(i)}\right),i\in\mathcal{I}_{\text{train}}\right\}\right)$
		\State Find available cases in training set and calibration set: $\tilde{\mathcal{I}}_{\text{train}}=\left\{i\in\mathcal{I}_{\text{train}}:M^{(i)}\preceq M^{(n+1)}\right\}$ and $\tilde{\mathcal{I}}=\left\{i\in[n]:M^{(i)}\preceq M^{(n+1)}\right\}$
		\State Compute the scores as $\tilde{V}^{(i)}=|Y^{(i)}-\hat{\mu}\circ\hat{\Phi}\left(X^{(i)},M^{(n+1)}\right)|,i\in\tilde{\mathcal{I}}_{\text{train}}\cup\tilde{\mathcal{I}}$
		\State Compute $V_{\alpha,h}^{(i)}$ for $i\in\tilde{\mathcal{I}}$ based on (\ref{localized score function}) as $V_{\alpha,h}^{(i)}=\tilde{V}^{(i)}-Q_{1-\alpha}\left(\hat{F}_{h}^{(i)}\right)$
		\State \textbf{Output}: let $d=Q_{1-\alpha}\left(\hat{F}_{h}^{(n+1)}\right)+Q_{1-\alpha}\left(\frac{1}{|\tilde{\mathcal{I}}|+1}\sum_{i\in\tilde{\mathcal{I}}}\delta_{V_{\alpha,h}^{(i)}}+\frac{1}{|\tilde{\mathcal{I}}|+1}\delta_{+\infty}\right)$,
		\begin{equation*}
			\hat{\mathcal{C}}_{n}\left(X^{(n+1)},M^{(n+1)}\right)=\left[\hat{\mu}\circ\hat{\Phi}\left(X^{(n+1)},M^{(n+1)}\right)-d,\hat{\mu}\circ\hat{\Phi}\left(X^{(n+1)},M^{(n+1)}\right)+d\right].
		\end{equation*}
	\end{algorithmic}
	\label{algorithm:lcp}
\end{algorithm}
\begin{remark}
	In Algorithm \ref{algorithm:lcp}, Lines 3-4 learn the imputation and regression functions, respectively. Line 5 finds the available cases for the missing pattern of the test point $M^{(n+1)}$ in the training and calibration set. Line 6 computes the original scores. Line 7 computes the new scores. 
\end{remark}

\begin{assumption}\label{holder_clsss_condition}
	Assume that $p(x)$, the probability density function of $X\in\mathbb{R}^{d}$, belongs to Holder class $\Sigma(\gamma,L)$. That is $p\in\Sigma(\gamma,L)$.
\end{assumption}

Assumption \ref{holder_clsss_condition} mainly restricts the space of probability density function of $X\in\mathbb{R}^{d}$ for establishing uniform consistency of the nonparametric estimation of conditional distribution  $\hat{F}_{h}(V=v\mid X=x,M=M^{(n+1)})$ to the true conditional distribution  $\hat{F}(V=v\mid X=x,M=M^{(n+1)})$, so that the localized conformal interval can achieve asymptotic conditional coverage guarantee.

\begin{theorem}\label{theorem:LCP}
	Given i.i.d. observations $(X^{(i)},M^{(i)},Y^{(i)}), i\in [n]$, and suppose Assumptions \ref{symmetry}-\ref{general assumption} hold, and the missing data mechanism is MCAR. For a new observation $(X^{(n+)},M^{(n+1)})$, the localized conformal prediction interval defined in Algorithm \ref{algorithm:lcp} is marginally valid,
	\begin{equation*}
		\mathbb{P}\left\{Y^{(n+1)}\in\hat{\mathcal{C}}_{n}\left(X^{(n+1)},M^{(n+1)}\right)\right\}\geq 1-\alpha.
	\end{equation*}
	Moreover, it is also mask-conditionally valid, i.e., for $m\in\{0,1\}^{d}$,
	\begin{equation*}
		\mathbb{P}\left\{Y^{(n+1)}\in\hat{\mathcal{C}}_{n}\left(X^{(n+1)},m\right)\mid M^{(n+1)}=m\right\}\geq 1-\alpha.
	\end{equation*}
	Furthermore, if Assumption \ref{holder_clsss_condition} is also satisfied, the localized conformal prediction is asymptotically conditional valid in the sense that, for $x\in\mathbb{R}^{d}$ and $m\in\{0,1\}^{d}$,
	\begin{equation*}
		\mathbb{P}\left\{Y^{(n+1)}\in\hat{\mathcal{C}}_{n}\left(x,m\right)\mid X^{(n+1)}=x,M^{(n+1)}=m\right\}\geq 1-\alpha,
	\end{equation*}
	as $n\to\infty$.
	
\end{theorem}

\begin{remark}
	Theorem \ref{theorem:LCP} demonstrates that, under certain conditions, the proposed localized conformal prediction approach is marginally valid, mask-conditionally valid and conditionally valid. The kernel function used in this framework can be any standard kernel, such as the Gaussian kernel. To measure the distance between covariates, we can employ the Heterogeneous Euclidean-Overlap Metric (HEOM) \citep{wilson1997improved}. For bandwidth selection, we follow the approach of \cite{han2022split}, utilizing the median of all pairwise distances between distinct data points within the training and calibration set.

\end{remark}

\begin{remark}
	
	We theoretically demonstrate that localized conformal prediction (LCP) asymptotically achieves conditional coverage on covariates, meaning that LCP approaches the oracle interval with minimal interval length \citep{lei2018distribution}. In contrast, nonexchangeable conformal prediction does not possess such asymptotic conditional coverage. Consequently, the prediction intervals produced by LCP tend to be shorter than those of nonexchangeable conformal prediction. 
	This finding aligns with the intuition that localized conformal prediction leverages more information about the distance between covariates compared to nonexchangeable conformal prediction. Our claim is further supported by both simulation studies and real data analyses, reinforcing the advantages of LCP in providing efficient prediction intervals.
	
\end{remark}

\section{Simulation studies}\label{Simulation studies}
In this section, we evaluate the empirical performance of the proposed nonexchangeable conformal prediction (nexCP) and localized conformal prediction (LCP). We compare the performance of nexCP and LCP against CQR-MDA-Exact, CQR-MDA-Nested \citep{zaffran2023conformal}, CQR \citep{romano2019conformalized}, and CP as defined in (\ref{naive conformal}). In all experiments, data are imputed using Multivariate Imputation by Chained Equations (MICE) \citep{pedregosa2011scikit}. Predictive regression models are then fitted on the imputed data. For CQR-MDA-Exact, CQR-MDA-Nested, and CQR, the predictive regression algorithm is a neural network trained to minimize quantile loss, with the network architecture specified in \cite{sesia2021conformal}. Similarly, for CP, nexCP, and LCP, the regression model follows the same architecture but optimizes for the least square loss. Evaluation metrics include marginal coverage (\ref{marginal validity}), mask-conditional coverage (\ref{mask-conditional validity}), and prediction interval lengths. The miscoverage rate is set to $\alpha = 0.1$.

\subsection{Synthetic experiments}

In synthetic experiments, We adopt a simulation design similar to that in \cite{zaffran2023conformal}, to enable direct comparison. Data are generated following the Gaussian linear model:
\begin{equation}
	Y=\beta^{T}X+\varepsilon.
\end{equation}
 Specifically, the dimension of $X$ is set to 3, 5, and 8, respectively. The covariates are drawn from a normal distribution, $X \sim N(\mu, \Sigma)$, where $\mu = (1, \dots, 1)^{T}$ and $\Sigma = \phi (1, \dots, 1)^{T} (1, \dots, 1) + (1 - \phi)I_{d}$, with $\phi = 0.8$. Gaussian noise is added, with $\varepsilon \sim N(0,1)$. The regression coefficient is set to $\beta = (1, 2, -1, 3, -0.5, -1, 0.3, 1.7)^{T}$. For dimensions $d=3$ and $d=5$, only the first $d$ covariates and corresponding parameters are used in the model.

We consider different missing data mechanisms, including MCAR, MAR, and MNAR. For the MCAR mechanism, missingness is independent of $X$, and all covariates may be missing. In the MAR setting, for $d=3$, the first component of $X$ is fully observed, while the last two components may be missing; for $d=5$, the last two components may be missing, and for $d=8$, the last three components may be missing. For the MNAR mechanism, we employ a self-masked mechanism with $\mathbb{P}\{M|X\}={\textstyle\prod_{k\in [d]}}\mathbb{P}\{M_{k}|X_{k}\}$, as described by \citep{le2020neumiss}, where all covariates may be missing. In all scenarios, the missing rate is set to $20\%$ for each missing covariate.

The training and calibration sets consist of 500 and 250 samples, respectively. For evaluating marginal coverage and interval length, the test set size is 2000. To assess the mask-conditional coverage,  $\mathbb{P}\left\{Y\in \hat{C}_{\alpha}\left(X,m\right)\mid M=m\right\}$, and the interval length for each $ m\in\{0,1\}^{d}$, we maintain a fixed number of observations per mask. This approach controls for variability in coverage and prediction interval length unrelated to $\mathbb{P}\{M = m\}$. As $d$ increases, the number of possible missing data patterns grows exponentially; for example, with $d = 8$, there are $2^{8} - 1 = 255$ unique patterns. When $d = 5$ or $d = 8$ under MCAR and MNAR settings, we present coverage and interval length as a function of mask size, serving as a proxy for mask-conditional coverage. For each mask size, 100 observations are drawn according to the distribution of $M|\text{size}(M)$ in the test set. This evaluation is based on 50 replications, with the coverage and interval lengths averaged across experiments.


%
Tables \ref{table:d=3}-\ref{table:d=8} present the simulation results. Overall, all six methods achieve marginal validity. However, CQR-MDA-Nested appears conservative, with coverage levels significantly exceeding the nominal coverage level of $1 - \alpha = 0.9$. 
In terms of mask-conditional validity, CP and CQR do not meet this criterion. Specifically, CP?s prediction interval length remains invariant to the mask, and while CQR?s prediction interval length varies slightly with the mask, this variation is insufficient to achieve mask-conditional validity. The remaining methods, CQR-MDA-Exact, CQR-MDA-Nested, nexCP, and LCP, are mask-conditionally valid, with CQR-MDA-Nested again demonstrating conservativeness.
Furthermore, the proposed nexCP and LCP are more efficient, yielding shorter prediction intervals than CQR-MDA-Exact and CQR-MDA-Nested. For instance, in Table \ref{table:d=5}, with the dimension of $X$ set to 5, both nexCP and LCP generally provide shorter intervals than CQR-MDA-Exact and CQR-MDA-Nested. Additionally, LCP shows slightly shorter intervals than nexCP, likely due to LCP?s ability to utilize quantitative information among covariates, whereas nexCP relies solely on qualitative information when constructing prediction intervals.

The distance metric introduced in nexCP and LCP allows the prediction interval lengths of these methods to remain relatively unaffected by changes in the missing data mechanism. For instance, in Table \ref{table:d=5} under the MCAR mechanism, the prediction interval lengths of CQR-MDA-Exact for both marginal and mask-conditional sets are as follows: 
\[
(\text{mar}, 0, 1, 2, 3, 4) = (5.508, 4.409, 5.476, 6.621, 7.956, 10.137).
\]
However, when the missing data mechanism shifts to MNAR, introducing a complex relationship between missingness and covariates, the interval lengths of CQR-MDA-Exact increase to 
\[
(\text{mar}, 0, 1, 2, 3, 4) = (5.769, 4.453, 5.660, 7.041, 8.732, 10.811).
\]
This lengthening reflects the increased complexity of missingness-covariate relationships under MNAR. In contrast, nexCP and LCP show minimal change in interval lengths with variations in the missing data mechanism. For example, under MCAR, the interval lengths for LCP are 
\[
(\text{mar}, 0, 1, 2, 3, 4) = (5.173, 3.885, 5.160, 6.426, 7.889, 10.257),
\]
and under MNAR, they are 
\[
(\text{mar}, 0, 1, 2, 3, 4) = (5.219, 3.892, 5.188, 6.462, 7.996, 10.321),
\]
exhibiting only slight increases. Thus, nexCP and LCP not only meet marginal and mask-conditional validity criteria but also demonstrate greater efficiency by maintaining shorter prediction interval lengths.

\begin{sidewaystable}
	\caption{Results for $d=3$ reported on test set of 50 repeated experiments with $\alpha = 0.1$.}\label{table:d=3}
	\begin{tabular*}{\textwidth}{@{\extracolsep\fill}lcccccccccccc}
		\toprule%
		& \multicolumn{2}{@{}c@{}}{CQR} & \multicolumn{2}{@{}c@{}}{CQR-MDA-Exact} & \multicolumn{2}{@{}c@{}}{CQR-MDA-Nested} & \multicolumn{2}{@{}c@{}}{CP}& \multicolumn{2}{@{}c@{}}{nexCP} & \multicolumn{2}{@{}c@{}}{LCP} \\\cmidrule{2-3}\cmidrule{4-5}\cmidrule{6-7}\cmidrule{8-9}\cmidrule{10-11}\cmidrule{12-13}%
	&	cov & len 	&	cov & len	&	cov & len	&	cov & len	&	cov & len	&	cov & len \\
		\midrule
		  &\multicolumn{12}{@{}c@{}}{MCAR}\\
		\cmidrule(lr){2-13}
		mar       & 0.898 & 4.341 & 0.900 & 4.366 & 0.987 & 6.938  & 0.894 & 4.209 & 0.904 & 4.321 & 0.899 & 4.256 \\
		{[}000{]} & 0.940 & 4.341 & 0.894 & 3.781 & 0.984 & 5.950  & 0.944 & 4.209 & 0.900 & 3.648 & 0.895 & 3.599 \\
		{[}001{]} & 0.908 & 4.328 & 0.890 & 4.098 & 0.987 & 6.349  & 0.908 & 4.209 & 0.899 & 4.118 & 0.895 & 4.063 \\
		{[}010{]} & 0.789 & 4.347 & 0.890 & 5.667 & 0.989 & 8.542  & 0.771 & 4.209 & 0.903 & 5.865 & 0.897 & 5.785 \\
		{[}011{]} & 0.789 & 4.320 & 0.895 & 5.536 & 0.997 & 9.867  & 0.772 & 4.209 & 0.904 & 5.850 & 0.900 & 5.742 \\
		{[}100{]} & 0.889 & 4.330 & 0.904 & 4.473 & 0.991 & 7.503  & 0.888 & 4.209 & 0.902 & 4.380 & 0.894 & 4.306 \\
		{[}101{]} & 0.886 & 4.352 & 0.897 & 4.481 & 0.995 & 9.055  & 0.888 & 4.209 & 0.905 & 4.448 & 0.899 & 4.374 \\
		{[}110{]} & 0.659 & 4.334 & 0.900 & 7.400 & 0.996 & 11.306 & 0.652 & 4.209 & 0.916 & 7.783 & 0.906 & 7.618 \\

		\midrule
		&\multicolumn{12}{c}{MAR}\\
		\cmidrule(lr){2-13}
		
		mar  & 0.903 & 4.220 & 0.908 & 4.319 & 0.978 & 5.959 & 0.903 & 4.105 & 0.907 & 4.147 & 0.904 & 4.108 \\
		{[}000{]} & 0.931 & 4.155 & 0.904 & 3.794 & 0.973 & 5.453 & 0.944 & 4.105 & 0.906 & 3.623 & 0.904 & 3.588 \\
		{[}001{]} & 0.897 & 4.239 & 0.904 & 4.368 & 0.980 & 5.926 & 0.901 & 4.105 & 0.911 & 4.254 & 0.905 & 4.182 \\
		{[}010{]} & 0.806 & 4.406 & 0.919 & 5.910 & 0.988 & 7.584 & 0.781 & 4.105 & 0.909 & 5.773 & 0.908 & 5.738 \\
		{[}011{]} & 0.814 & 4.458 & 0.922 & 5.921 & 0.987 & 7.632 & 0.778 & 4.105 & 0.910 & 5.813 & 0.908 & 5.718 \\
		
		\midrule
		&\multicolumn{12}{c}{MNAR}\\
		\cmidrule(lr){2-13}
		
		mar  & 0.899 & 4.446 & 0.901 & 4.480 & 0.988 & 7.094  & 0.901 & 4.330 & 0.900 & 4.304 & 0.896 & 4.244 \\
		{[}000{]} & 0.937 & 4.373 & 0.898 & 3.818 & 0.985 & 6.030  & 0.946 & 4.330 & 0.893 & 3.595 & 0.889 & 3.564 \\
		{[}001{]} & 0.917 & 4.454 & 0.905 & 4.273 & 0.989 & 6.550  & 0.917 & 4.330 & 0.905 & 4.140 & 0.898 & 4.053 \\
		{[}010{]} & 0.813 & 4.543 & 0.911 & 5.907 & 0.994 & 8.759  & 0.787 & 4.330 & 0.900 & 5.909 & 0.897 & 5.808 \\
		{[}011{]} & 0.810 & 4.595 & 0.909 & 6.002 & 0.997 & 10.168 & 0.788 & 4.330 & 0.906 & 5.901 & 0.899 & 5.792 \\
		{[}100{]} & 0.899 & 4.460 & 0.905 & 4.554 & 0.992 & 7.643  & 0.899 & 4.330 & 0.904 & 4.396 & 0.899 & 4.312 \\
		{[}101{]} & 0.898 & 4.501 & 0.913 & 4.669 & 0.996 & 9.539  & 0.898 & 4.330 & 0.911 & 4.534 & 0.902 & 4.440 \\
		{[}110{]} & 0.682 & 4.601 & 0.907 & 7.674 & 0.995 & 11.513 & 0.640 & 4.330 & 0.899 & 7.736 & 0.897 & 7.611 \\

		\botrule
	\end{tabular*}
	\footnotetext{Note: The results provide coverage and length of prediction intervals for each method across three different missing data mechanisms. Here, ``cov" and ``len" denote the average coverage and interval length, respectively, while ``mar" refers to ``marginal". Additionally, ``{[}000{]}" represents the mask $(0,0,0)$, which indicates that all covariates in the test set are observed. Other notations refer to specific masks.}
\end{sidewaystable}

\begin{sidewaystable}
	\caption{Results for $d=5$ reported on test set of 50 repeated experiments with $\alpha = 0.1$.}\label{table:d=5}
	\begin{tabular*}{\textwidth}{@{\extracolsep\fill}lcccccccccccc}
		\toprule%
		& \multicolumn{2}{@{}c@{}}{CQR} & \multicolumn{2}{@{}c@{}}{CQR-MDA-Exact} & \multicolumn{2}{@{}c@{}}{CQR-MDA-Nested} & \multicolumn{2}{@{}c@{}}{CP}& \multicolumn{2}{@{}c@{}}{nexCP} & \multicolumn{2}{@{}c@{}}{LCP} \\\cmidrule{2-3}\cmidrule{4-5}\cmidrule{6-7}\cmidrule{8-9}\cmidrule{10-11}\cmidrule{12-13}%
		&	cov & len 	&	cov & len	&	cov & len	&	cov & len	&	cov & len	&	cov & len \\
		\midrule
		&\multicolumn{12}{@{}c@{}}{MCAR}\\
		\cmidrule(lr){2-13}
	mar & 0.900 & 5.606 & 0.902 & 5.508  & 0.995 & 9.725  & 0.900 & 5.310 & 0.907 & 5.282  & 0.900 & 5.173  \\
	0        & 0.963 & 5.598 & 0.904 & 4.409  & 0.992 & 8.174  & 0.973 & 5.310 & 0.905 & 3.947  & 0.899 & 3.885  \\
	1        & 0.904 & 5.631 & 0.907 & 5.476  & 0.995 & 9.556  & 0.901 & 5.310 & 0.907 & 5.268  & 0.898 & 5.160  \\
	2        & 0.830 & 5.642 & 0.892 & 6.621  & 0.995 & 11.236 & 0.817 & 5.310 & 0.897 & 6.571  & 0.889 & 6.426  \\
	3        & 0.749 & 5.614 & 0.892 & 7.956  & 0.998 & 13.857 & 0.735 & 5.310 & 0.904 & 8.060  & 0.898 & 7.889  \\
	4        & 0.651 & 5.618 & 0.904 & 10.137 & 1.000 & 19.869 & 0.624 & 5.310 & 0.905 & 10.483 & 0.902 & 10.257 \\

		\midrule
		&\multicolumn{12}{c}{MAR}\\
		\cmidrule(lr){2-13}
		mar    & 0.903 & 4.760 & 0.903 & 4.827 & 0.981 & 7.160  & 0.903 & 4.519 & 0.905 & 4.467 & 0.903 & 4.420 \\
		{[}00000{]} & 0.944 & 4.648 & 0.900 & 4.060 & 0.980 & 6.413  & 0.956 & 4.519 & 0.904 & 3.715 & 0.902 & 3.689 \\
		{[}00001{]} & 0.933 & 4.777 & 0.903 & 4.283 & 0.981 & 6.623  & 0.944 & 4.519 & 0.905 & 3.876 & 0.901 & 3.824 \\
		{[}00010{]} & 0.753 & 5.104 & 0.928 & 7.718 & 0.991 & 9.946  & 0.707 & 4.519 & 0.916 & 7.339 & 0.910 & 7.245 \\
		{[}00011{]} & 0.767 & 5.191 & 0.919 & 7.681 & 0.993 & 10.086 & 0.707 & 4.519 & 0.912 & 7.439 & 0.905 & 7.281 \\

		\midrule
		&\multicolumn{12}{c}{MNAR}\\
		\cmidrule(lr){2-13}
		mar & 0.902 & 5.637 & 0.909 & 5.769  & 0.995 & 9.874  & 0.899 & 5.358 & 0.903 & 5.330  & 0.897 & 5.219  \\
		0        & 0.961 & 5.434 & 0.908 & 4.453  & 0.994 & 8.057  & 0.976 & 5.358 & 0.906 & 3.965  & 0.899 & 3.892  \\
		1        & 0.900 & 5.612 & 0.903 & 5.660  & 0.996 & 9.597  & 0.904 & 5.358 & 0.904 & 5.316  & 0.898 & 5.188  \\
		2        & 0.852 & 5.853 & 0.913 & 7.041  & 0.998 & 11.564 & 0.824 & 5.358 & 0.899 & 6.606  & 0.896 & 6.462  \\
		3        & 0.802 & 6.159 & 0.925 & 8.732  & 0.999 & 14.943 & 0.739 & 5.358 & 0.904 & 8.120  & 0.900 & 7.996  \\
		4        & 0.664 & 6.130 & 0.908 & 10.811 & 0.998 & 21.270 & 0.624 & 5.358 & 0.897 & 10.504 & 0.892 & 10.321 \\

		\botrule
	\end{tabular*}
	\footnotetext{Note:  ``cov" and ``len" denote the average coverage and length of prediction intervals, respectively. ``mar" stands for ``marginal", while ``{[}00000{]}" indicates the mask $(0,0,0,0,0)$, where all covariates are observed. Other masks are specified similarly. ``0-4" represents the pattern size; for example, a pattern size of 1 indicates one missing covariate, which may occur in any covariate component.}
\end{sidewaystable}

\begin{sidewaystable}
	\caption{Results for $d=8$ reported on test set of 50 repeated experiments with $\alpha = 0.1$.}\label{table:d=8}
	\begin{tabular*}{\textwidth}{@{\extracolsep\fill}lcccccccccccc}
		\toprule%
		& \multicolumn{2}{@{}c@{}}{CQR} & \multicolumn{2}{@{}c@{}}{CQR-MDA-Exact} & \multicolumn{2}{@{}c@{}}{CQR-MDA-Nested} & \multicolumn{2}{@{}c@{}}{CP}& \multicolumn{2}{@{}c@{}}{nexCP} & \multicolumn{2}{@{}c@{}}{LCP} \\\cmidrule{2-3}\cmidrule{4-5}\cmidrule{6-7}\cmidrule{8-9}\cmidrule{10-11}\cmidrule{12-13}%
		&	cov & len 	&	cov & len	&	cov & len	&	cov & len	&	cov & len	&	cov & len \\
		\midrule
		&\multicolumn{12}{@{}c@{}}{MCAR}\\
		\cmidrule(lr){2-13}

		mar & 0.904 & 6.084 & 0.904 & 6.044  & 0.996 & 10.250 & 0.904 & 5.799 & 0.913 & 5.902  & 0.904 & 5.726  \\
		0   & 0.975 & 6.075 & 0.911 & 4.764  & 0.995 & 8.775  & 0.976 & 5.799 & 0.915 & 4.395  & 0.909 & 4.285  \\
		1   & 0.928 & 6.084 & 0.901 & 5.545  & 0.995 & 9.645  & 0.933 & 5.799 & 0.908 & 5.403  & 0.901 & 5.225  \\
		2   & 0.890 & 6.072 & 0.906 & 6.330  & 0.997 & 10.567 & 0.894 & 5.799 & 0.917 & 6.303  & 0.906 & 6.100  \\
		3   & 0.846 & 6.097 & 0.905 & 7.117  & 0.996 & 11.575 & 0.835 & 5.799 & 0.902 & 7.121  & 0.895 & 6.926  \\
		4   & 0.802 & 6.073 & 0.904 & 7.901  & 0.998 & 12.813 & 0.796 & 5.799 & 0.915 & 8.031  & 0.909 & 7.800  \\
		5   & 0.750 & 6.123 & 0.899 & 8.903  & 0.998 & 14.506 & 0.723 & 5.799 & 0.908 & 9.062  & 0.903 & 8.837  \\
		6   & 0.696 & 6.100 & 0.904 & 10.141 & 0.999 & 17.321 & 0.667 & 5.799 & 0.910 & 10.404 & 0.901 & 10.144 \\
		7   & 0.591 & 6.060 & 0.910 & 12.599 & 1.000 & 23.405 & 0.570 & 5.799 & 0.922 & 12.951 & 0.912 & 12.659 \\
		
		\midrule
		&\multicolumn{12}{c}{MAR}\\
		\cmidrule(lr){2-13}
mar               & 0.901 & 4.288 & 0.911 & 4.423 & 0.975 & 5.831 & 0.897 & 3.978 & 0.906 & 4.084 & 0.902 & 4.028 \\
{[}000{]}    & 0.925 & 4.185 & 0.911 & 4.036 & 0.973 & 5.384 & 0.930 & 3.978 & 0.909 & 3.733 & 0.908 & 3.697 \\
{[}001{]} & 0.841 & 4.353 & 0.916 & 5.329 & 0.984 & 6.824 & 0.822 & 3.978 & 0.910 & 5.036 & 0.904 & 4.953 \\
{[}010{]} & 0.921 & 4.276 & 0.911 & 4.122 & 0.975 & 5.555 & 0.924 & 3.978 & 0.908 & 3.768 & 0.905 & 3.725 \\
{[}011{]} & 0.833 & 4.473 & 0.916 & 5.517 & 0.983 & 7.068 & 0.813 & 3.978 & 0.910 & 5.174 & 0.904 & 5.055 \\
{[}100{]} & 0.896 & 4.411 & 0.906 & 4.560 & 0.980 & 6.022 & 0.883 & 3.978 & 0.897 & 4.163 & 0.893 & 4.097 \\
{[}101{]} & 0.834 & 4.528 & 0.909 & 5.581 & 0.988 & 7.232 & 0.793 & 3.978 & 0.908 & 5.271 & 0.902 & 5.153 \\
{[}110{]} & 0.896 & 4.379 & 0.908 & 4.545 & 0.980 & 6.041 & 0.889 & 3.978 & 0.906 & 4.197 & 0.903 & 4.118 \\
{[}111{]} & 0.822 & 4.529 & 0.911 & 5.667 & 0.986 & 7.325 & 0.792 & 3.978 & 0.909 & 5.336 & 0.902 & 5.218 \\

		\midrule
		&\multicolumn{12}{c}{MNAR}\\
		\cmidrule(lr){2-13}

		mar & 0.904 & 6.025 & 0.909 & 6.171  & 0.996 & 10.286 & 0.903 & 5.802 & 0.911 & 5.873  & 0.901 & 5.689  \\
		0   & 0.963 & 5.762 & 0.895 & 4.606  & 0.995 & 8.499  & 0.974 & 5.802 & 0.910 & 4.376  & 0.900 & 4.249  \\
		1   & 0.926 & 5.888 & 0.913 & 5.576  & 0.995 & 9.501  & 0.929 & 5.802 & 0.909 & 5.362  & 0.900 & 5.172  \\
		2   & 0.883 & 6.022 & 0.909 & 6.501  & 0.996 & 10.609 & 0.890 & 5.802 & 0.914 & 6.269  & 0.900 & 6.054  \\
		3   & 0.859 & 6.284 & 0.921 & 7.538  & 0.996 & 11.920 & 0.835 & 5.802 & 0.908 & 7.103  & 0.901 & 6.902  \\
		4   & 0.834 & 6.668 & 0.922 & 8.622  & 0.999 & 13.443 & 0.789 & 5.802 & 0.913 & 7.900  & 0.908 & 7.730  \\
		5   & 0.789 & 6.845 & 0.921 & 9.757  & 0.998 & 15.538 & 0.723 & 5.802 & 0.901 & 8.938  & 0.896 & 8.787  \\
		6   & 0.729 & 6.853 & 0.915 & 11.048 & 0.999 & 18.962 & 0.666 & 5.802 & 0.902 & 10.364 & 0.894 & 10.159 \\
		7   & 0.589 & 6.517 & 0.904 & 13.172 & 1.000 & 25.192 & 0.538 & 5.802 & 0.897 & 12.935 & 0.894 & 12.696 \\

		\botrule
	\end{tabular*}
	\footnotetext{Note: ``cov" and ``len" denote the average coverage and length of prediction intervals, respectively. ``mar" stands for ``marginal". For simplicity, we only display the last three components of the mask; for example, ``{[}000{]}" indicates the mask $(0,0,0,0,0,0,0,0)$, meaning all covariates are observed. Other masks are specified similarly. ``0-7" indicates the pattern size; for instance, a pattern size of 1 signifies one missing covariate, which may occur in any covariate component.}
\end{sidewaystable}

\subsection{Semi-synthetic experiments}


In this subsection, we conduct semi-synthetic experiments to evaluate the performance of existing methods. Concrete is a crucial material in civil engineering, and its compressive strength is a highly nonlinear function of age and composition. We utilize the \textit{Concrete} dataset from the UCI Machine Learning Repository \citep{dua2017uci}. This dataset comprises 1,030 samples with eight continuous covariates and one response variable. We introduce missing values in the dataset in accordance with the synthetic experiments, employing mechanisms such as MCAR, MAR, and MNAR. 
The training and calibration sets are set to 330 and 100 samples, respectively, while the marginal test set contains 200 samples, and the mask-conditional test set has 100 samples. It is important to note that due to the relatively small sample size, some patterns may appear with low or null frequency. For the MCAR and MNAR mechanisms, we report only scenarios with a pattern size of four or fewer. In the MAR mechanism, the last three covariate components are subject to missingness. The experiments are repeated 100 times.

Table \ref{table:concrete} summarizes the simulation results, which are consistent with those from the synthetic experiments. The proposed methods, nexCP and LCP, are not only marginally valid but also mask-conditionally valid. Furthermore, they demonstrate greater efficiency than existing methods. In particular, in the \textit{Concrete} dataset, the prediction interval length of LCP is shorter than that of nexCP, establishing LCP as the most efficient solution.

\begin{sidewaystable}
	\caption{Results for ``Concrete" dataset of 100 repeated experiments with $\alpha = 0.1$.}\label{table:concrete}
	\begin{tabular*}{\textwidth}{@{\extracolsep\fill}lcccccccccccc}
		\toprule%
		& \multicolumn{2}{@{}c@{}}{CQR} & \multicolumn{2}{@{}c@{}}{CQR-MDA-Exact} & \multicolumn{2}{@{}c@{}}{CQR-MDA-Nested} & \multicolumn{2}{@{}c@{}}{CP}& \multicolumn{2}{@{}c@{}}{nexCP} & \multicolumn{2}{@{}c@{}}{LCP} \\\cmidrule{2-3}\cmidrule{4-5}\cmidrule{6-7}\cmidrule{8-9}\cmidrule{10-11}\cmidrule{12-13}%
		&	cov & len 	&	cov & len	&	cov & len	&	cov & len	&	cov & len	&	cov & len \\
		\midrule
		&\multicolumn{12}{@{}c@{}}{MCAR}\\
		\cmidrule(lr){2-13}
		mar & 0.903 & 43.954 & 0.903 & 45.512 & 0.992 & 65.017 & 0.898 & 39.640 & 0.918 & 44.442 & 0.901 & 40.939 \\
		0   & 0.943 & 43.890 & 0.904 & 42.604 & 0.985 & 56.232 & 0.952 & 39.640 & 0.919 & 38.575 & 0.902 & 35.373 \\
		1   & 0.922 & 43.787 & 0.900 & 43.022 & 0.990 & 61.697 & 0.925 & 39.640 & 0.922 & 41.409 & 0.899 & 37.348 \\
		2   & 0.896 & 44.033 & 0.902 & 45.844 & 0.993 & 67.393 & 0.890 & 39.640 & 0.923 & 45.526 & 0.905 & 42.080 \\
		3   & 0.861 & 43.863 & 0.903 & 49.449 & 0.994 & 72.498 & 0.847 & 39.640 & 0.921 & 50.800 & 0.905 & 47.636 \\
		4   & 0.821 & 44.002 & 0.899 & 53.385 & 0.996 & 77.250 & 0.790 & 39.640 & 0.915 & 55.462 & 0.900 & 52.933 \\

		\midrule
		&\multicolumn{12}{c}{MAR}\\
		\cmidrule(lr){2-13}
mar       & 0.899 & 39.515 & 0.905 & 40.185 & 0.969 & 49.543 & 0.903 & 35.279 & 0.909 & 35.781 & 0.900 & 34.696 \\
{[}000{]} & 0.918 & 38.861 & 0.904 & 37.807 & 0.962 & 46.694 & 0.936 & 35.279 & 0.908 & 31.900 & 0.900 & 30.905 \\
{[}001{]} & 0.838 & 40.560 & 0.906 & 47.140 & 0.984 & 57.262 & 0.770 & 35.279 & 0.917 & 50.769 & 0.909 & 49.366 \\
{[}010{]} & 0.921 & 39.293 & 0.911 & 38.513 & 0.972 & 47.751 & 0.932 & 35.279 & 0.911 & 32.251 & 0.900 & 31.181 \\
{[}011{]} & 0.826 & 40.951 & 0.910 & 47.894 & 0.983 & 58.147 & 0.747 & 35.279 & 0.902 & 50.479 & 0.895 & 49.152 \\
{[}100{]} & 0.907 & 39.491 & 0.905 & 39.524 & 0.969 & 49.002 & 0.931 & 35.279 & 0.913 & 32.914 & 0.902 & 31.734 \\
{[}101{]} & 0.842 & 40.660 & 0.921 & 48.074 & 0.988 & 58.412 & 0.793 & 35.279 & 0.923 & 50.696 & 0.913 & 49.367 \\
{[}110{]} & 0.912 & 40.941 & 0.912 & 40.928 & 0.972 & 51.162 & 0.924 & 35.279 & 0.904 & 33.309 & 0.896 & 32.375 \\
{[}111{]} & 0.847 & 42.239 & 0.921 & 49.312 & 0.989 & 60.228 & 0.765 & 35.231 & 0.919 & 50.410 & 0.912 & 49.309 \\

		\midrule
		&\multicolumn{12}{c}{MNAR}\\
		\cmidrule(lr){2-13}
		mar & 0.910 & 43.905 & 0.908 & 44.848 & 0.993 & 64.944 & 0.905 & 39.703 & 0.929 & 44.949 & 0.911 & 41.303 \\
		0   & 0.948 & 43.528 & 0.910 & 40.987 & 0.989 & 55.902 & 0.957 & 39.703 & 0.936 & 39.964 & 0.920 & 35.983 \\
		1   & 0.929 & 43.634 & 0.906 & 42.479 & 0.993 & 61.438 & 0.930 & 39.703 & 0.930 & 42.431 & 0.913 & 38.112 \\
		2   & 0.899 & 43.972 & 0.907 & 45.571 & 0.993 & 67.368 & 0.893 & 39.703 & 0.923 & 45.744 & 0.908 & 42.406 \\
		3   & 0.864 & 44.384 & 0.907 & 49.688 & 0.995 & 73.219 & 0.843 & 39.703 & 0.922 & 50.482 & 0.907 & 47.608 \\
		4   & 0.829 & 44.717 & 0.906 & 54.172 & 0.995 & 78.627 & 0.789 & 39.703 & 0.915 & 55.879 & 0.902 & 53.246 \\

		\botrule
	\end{tabular*}
	\footnotetext{Note: ``cov" and ``len" denote the average coverage and length of prediction intervals, respectively. ``mar" stands for ``marginal". For simplicity, we only display the last three components of the mask; for example, ``{[}000{]}" indicates the mask $(0,0,0,0,0,0,0,0)$, meaning all covariates are observed. Other masks are specified similarly. ``0-4" indicates the pattern size; for instance, a pattern size of 1 signifies one missing covariate, which may occur in any covariate component.}
\end{sidewaystable}

\section{Real data analysis}\label{Real data analysis}

In this section, we analyze the Chinese Provincial Legal Funding Dataset (CPLFD) from the Harvard Dataverse \citep{wang2015tying}, which has also been examined in the statistical literature, such as \citep{zhan2023partial,liu2024penalized}. The dataset includes key economic indicators at the provincial level for the period between 1995 and 2006.

\begin{sidewaystable}
	\caption{Results for CPLFD of 100 repeated experiments with $\alpha = 0.1$.}\label{table:CPLFD}
	\begin{tabular*}{\textwidth}{@{\extracolsep\fill}lcccccccccccc}
		\toprule%
		& \multicolumn{2}{@{}c@{}}{CQR} & \multicolumn{2}{@{}c@{}}{CQR-MDA-Exact} & \multicolumn{2}{@{}c@{}}{CQR-MDA-Nested} & \multicolumn{2}{@{}c@{}}{CP}& \multicolumn{2}{@{}c@{}}{nexCP} & \multicolumn{2}{@{}c@{}}{LCP} \\\cmidrule{2-3}\cmidrule{4-5}\cmidrule{6-7}\cmidrule{8-9}\cmidrule{10-11}\cmidrule{12-13}%
		&	cov & len 	&	cov & len	&	cov & len	&	cov & len	&	cov & len	&	cov & len \\
		\midrule
mar       & 0.900 & 3.789 & 0.907 & 3.876 & 0.981 & 6.207 & 0.902 & 3.575 & 0.907 & 3.719 & 0.896 & 3.502 \\
{[}000{]} & 0.917 & 3.592 & 0.905 & 3.593 & 0.974 & 5.024 & 0.949 & 3.575 & 0.916 & 3.294 & 0.901 & 3.031 \\
{[}010{]} & 0.766 & 3.866 & 0.915 & 5.527 & 0.999 & 8.685 & 0.756 & 3.563 & 0.914 & 5.898 & 0.899 & 5.494 \\
{[}001{]} & 0.930 & 3.849 & 0.906 & 3.550 & 0.979 & 6.027 & 0.922 & 3.579 & 0.902 & 3.311 & 0.893 & 3.161 \\

		\botrule
	\end{tabular*}
	\footnotetext{Note: ``cov" and ``len" denote the average coverage and length of prediction intervals, respectively. ``mar" stands for ``marginal". Additionally, ``{[}000{]}" represents the mask $(0,0,0)$, which indicates that all covariates in the test set are observed. Other notations refer to specific masks.}
\end{sidewaystable}

\begin{figure}[htb]
	\centering
	\begin{tabular}{cccc}
		\includegraphics[width=0.2\textwidth]{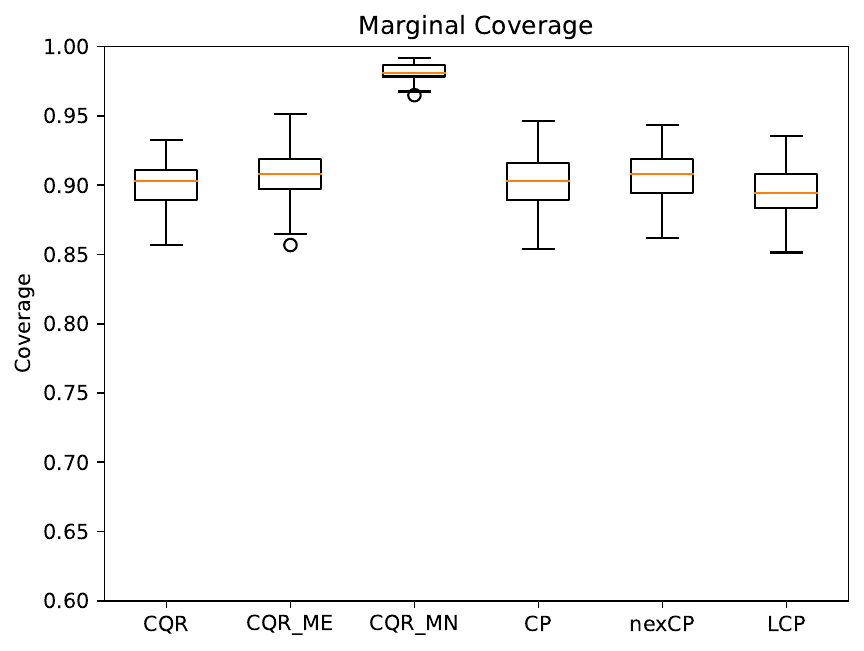}&
		\includegraphics[width=0.2\textwidth]{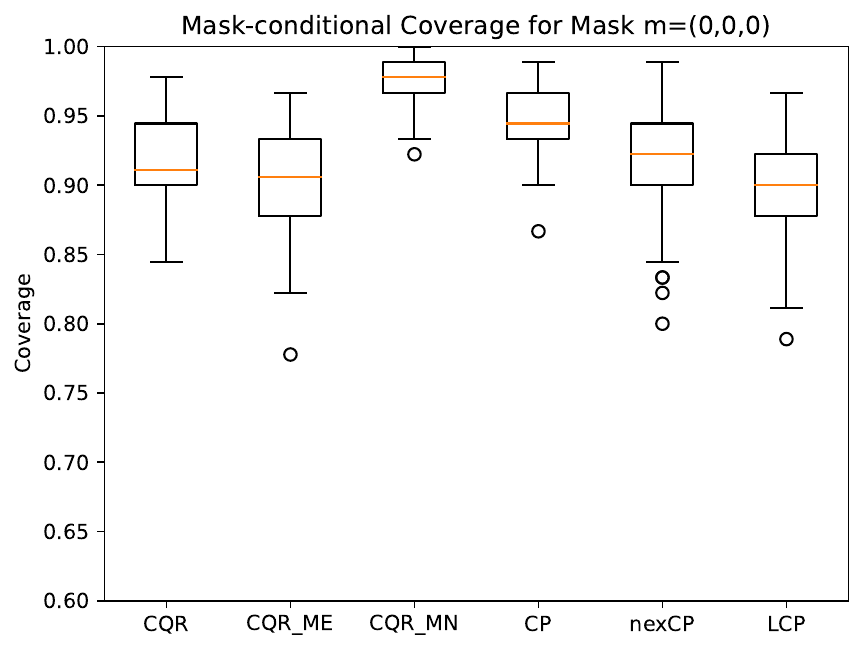}&
		\includegraphics[width=0.2\textwidth]{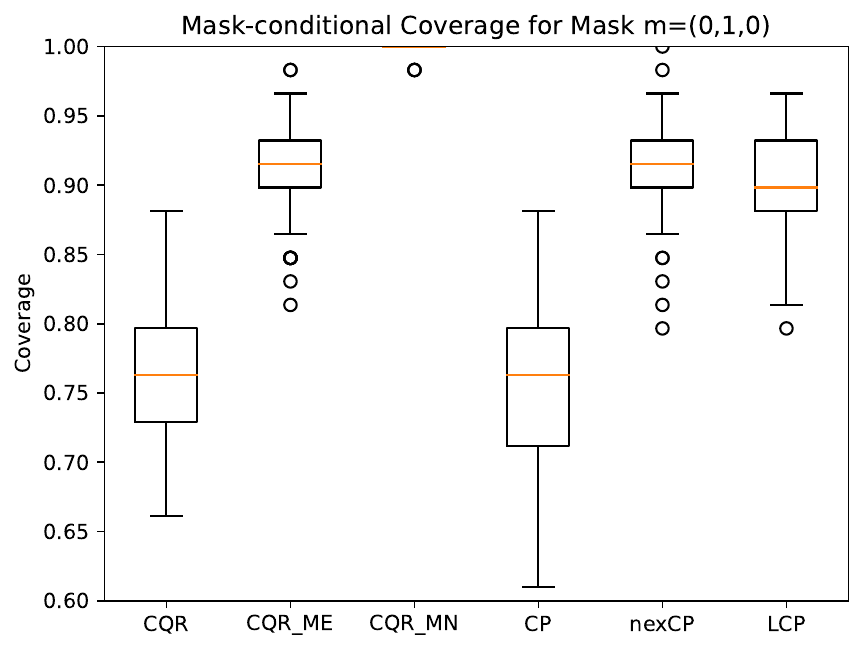}&
		\includegraphics[width=0.2\textwidth]{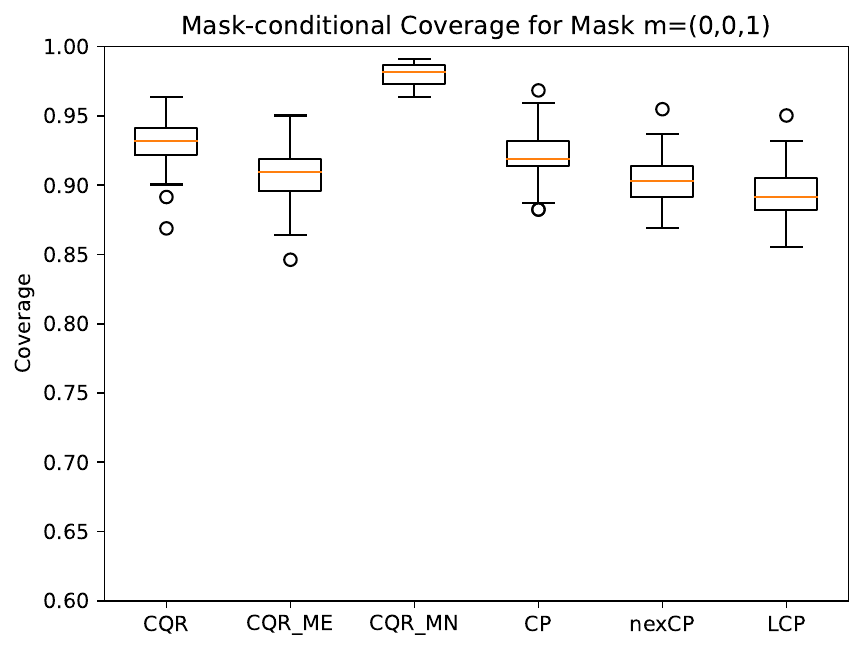} \\
		
		\includegraphics[width=0.2\textwidth]{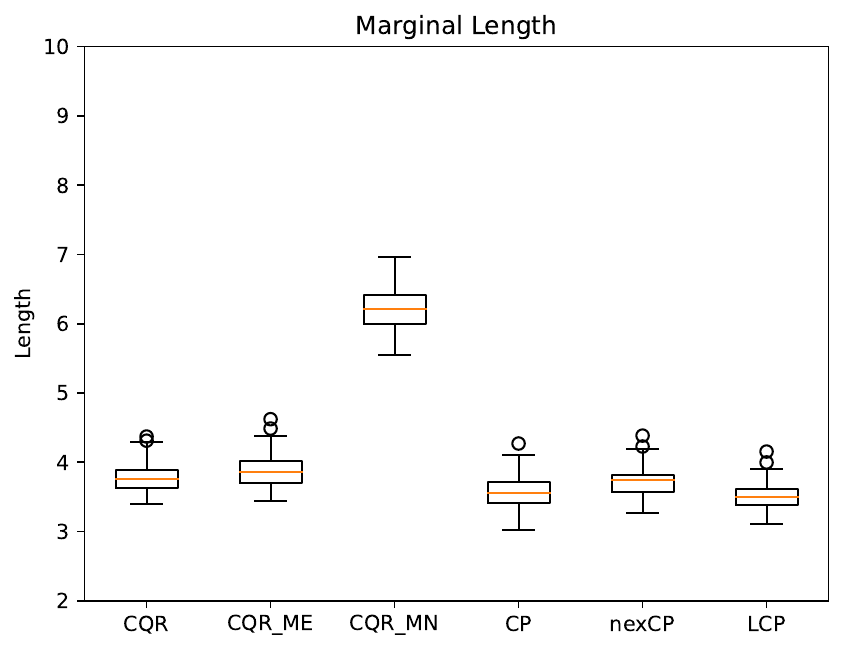}&
		\includegraphics[width=0.2\textwidth]{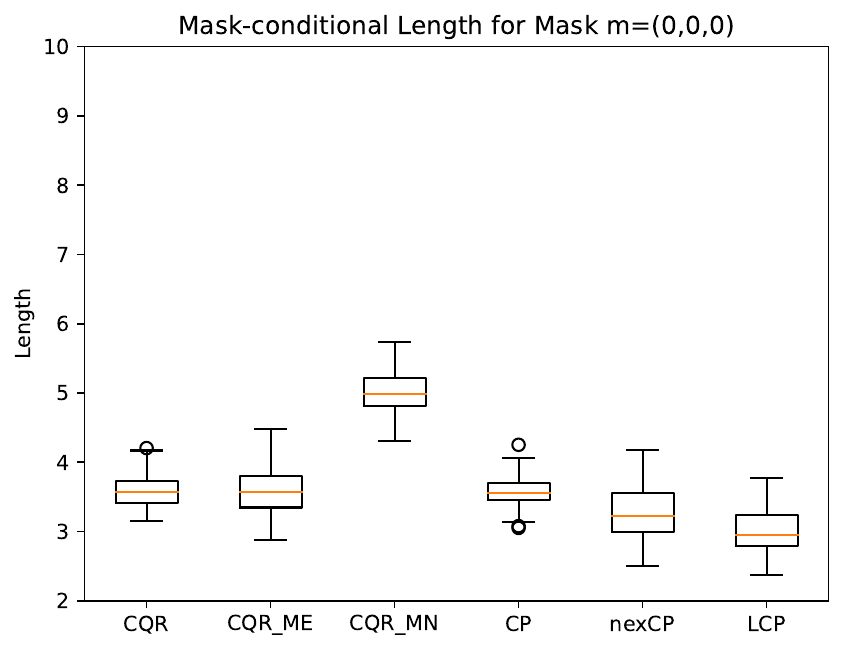}&
		\includegraphics[width=0.2\textwidth]{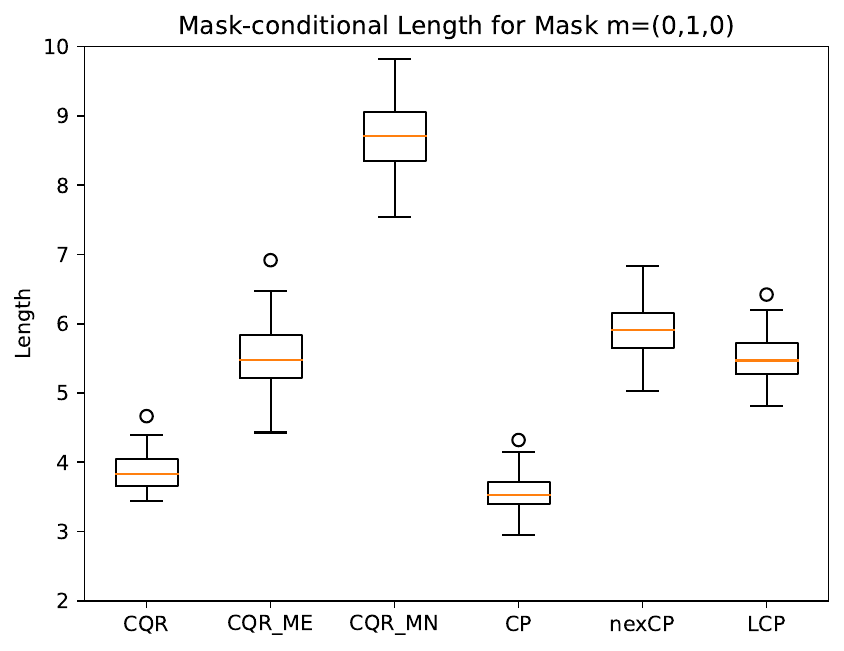}&
		\includegraphics[width=0.2\textwidth]{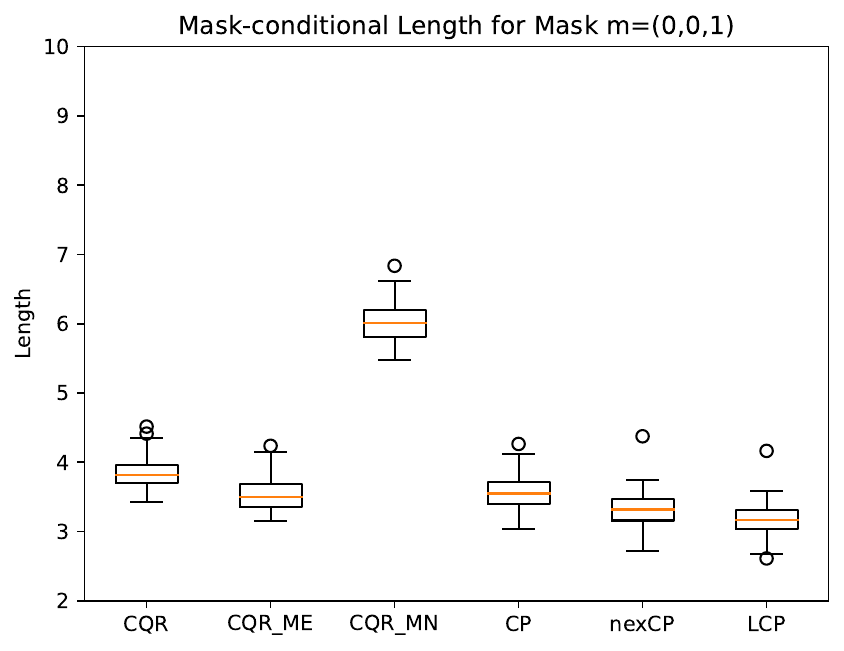}      
	\end{tabular}
	\caption{Box plot of the results for CPLFD. CQR\_ME denotes CQR-MDA-Exact; CQR\_MN denotes CQR-MDA-Nested. From the left to the right, each column corresponds to marginal coverage and length, mask-conditional coverage and length, i.e., (0,0,0), (0,1,0) and (0,0,1), respectively.}
	\label{figure:CPLFD}
\end{figure}

The dataset consists of 372 records and 60 variables. We aim to predict \textit{Foreign Direct Investment (FDI)} using three covariates: \textit{1st Industry}, \textit{Utilized Foreign Capital}, and \textit{Court Income}. The response variable, FDI, is fully observed but substantially skewed, so a log transformation is applied. After the transformation, the dataset includes 90 complete cases with mask $m=(0,0,0)$, 59 cases with mask $m=(0,1,0)$, and 221 cases with mask $m=(0,0,1)$. Observations with other values of $m$ are sparse and thus excluded. Due to the small sample size, we adopt a strategy similar to that used by \cite{zaffran2023conformal}. For each data point, only one prediction is made to avoid excessive repetitions. The data set is divided into 5 folds, with each fold predicted by running conformal procedures on the remaining 4. This process is repeated 100 times at a significance level of $\alpha = 0.1$.

Table \ref{table:CPLFD} presents the average coverage and length for different methods with respect to marginal and mask-conditional validity, while Figure \ref{figure:CPLFD} displays a box plot of the results. All six methods are marginally valid; however, only CQR-MDA-Exact, CQR-MDA-Nested, nexCP, and LCP demonstrate mask-conditional validity, with CQR-MDA-Nested being overly conservative. In addition, the proposed methods, nexCP and LCP, are more efficient, achieving smaller prediction interval lengths while maintaining the coverage rate. 

Interestingly, despite using a neural network as the regression function, we are still able to assess variable importance when mask-conditional validity is satisfied. For example, consider the LCP method in Table \ref{table:CPLFD}. When all three covariates are fully observed (i.e., for mask $m=(0,0,0)$), the average prediction interval length is 3.031. However, when one covariate, Utilized Foreign Capital, is missing (mask $m=(0,1,0)$), the average prediction interval length increases to 5.494, much longer than that for the fully observed covariates. This indicates that Utilized Foreign Capital plays a crucial role in predicting the response variable, FDI. On the other hand, when Court Income is missing (mask $m=(0,0,1)$), the average prediction interval length is 3.161, which is similar to the fully observed case. This suggests that Court Income is not so important for predicting FDI. It is important to note that these conclusions hold under the condition of mask-conditional validity. For instance, if we consider CP or CQR, these findings do not emerge, highlighting the significance of mask-conditional validity.

\section{Conclusion}\label{Conclusion}
As statistics and machine learning algorithms are increasingly applied in high-stakes decision-making processes, it is essential to develop tools that quantify the uncertainty of predictions. Conformal prediction offers a prediction set with a finite-sample coverage guarantee for any underlying predictive algorithm under mild conditions, making it applicable to a wide range of applications. However, as the volume of data increases, so does the amount of missing data, further complicating uncertainty quantification.

In this article, we investigate the problem of conformal prediction where missing values may arise in the covariates of both training and testing data. We examine the impact of missing covariates on conformal prediction. Furthermore, we propose two conformal methods to handle missing covariates. Our extensive simulation studies, along with a real-data example, demonstrate the advantages of the proposed conformal prediction methods in the presence of missing covariates.

Throughout this article, we use the Heterogeneous Euclidean-Overlap Metric as the distance metric for the proposed methods. We also consider other distance functions, such as the Heterogeneous Value Difference Metric and the Mean Euclidean Distance \citep{santos2020distance}, but the results obtained with these alternative metrics are similar. The mask-conditional validity of the proposed methods requires the MCAR mechanism, which is a relatively strong assumption. This condition is also employed in \cite{zaffran2023conformal,zaffran2024predictive}. However, our simulation studies indicate that mask-conditional validity is maintained in all MCAR, MAR, and MNAR mechanisms. Therefore, it may be possible to relax the MCAR assumption when proving the mask-conditional validity of the proposed methods. Additionally, this article focuses solely on missingness in covariates; other forms of missingness, such as missingness in both covariates and the response, deserve further investigation.

\section*{Appendix}
\setcounter{equation}{0}
\setcounter{subsection}{0}
\renewcommand{\theequation}{A.\arabic{equation}}
\renewcommand{\thesubsection}{A.\arabic{subsection}}
\subsection{Proof of Theorem \ref{Marginal validity}.}
The proof of the marginal validity is quite standard, the similar strategy can also be found in, e.g., \cite{lei2018distribution,barber2023conformal,zaffran2023conformal}. By i.i.d. observations and Assumption \ref{symmetry}, the data points $(X^{(i)},M^{(i)},Y^{(i)}),i\in [n+1]$, are indeed exchangeable, and the regression and imputation algorithms treat these data points symmetrically. Thus, the absolute residuals $V^{(i)}=|Y^{(i)}-\hat{\mu}\circ\hat{\Phi}(X^{(i)},M^{(i)})|,i\in [n+1]$, are exchangeable.  

Next, we define the set of ``strange" points as
\begin{equation*}
	S(V)=\left\{i\in [n+1]: V^{(i)}>Q_{1-\alpha}\left(\sum_{j=1}^{n+1}\frac{1}{n+1}\cdot\delta_{V^{(j)}}\right)\right\}.
\end{equation*}
That is, an index $i$ corresponds to a ``strange" point if its residuals $V^{(i)}$ is one of the $\lfloor \alpha (n+1) \rfloor$ largest elements of $V^{(i)}$, for $i\in [n+1]$. By definition, this include at most $\alpha(n+1)$ entries, that is,
\begin{equation*}
	|S(V)|\leq\alpha(n+1).
\end{equation*}
By the definition of the split conformal prediction interval defined in (\ref{naive conformal}), we can see that $Y^{(n+1)}\notin\hat{\mathcal{C}}_{n}\left(X^{(n+1)},M^{(n+1)}\right)$ (i.e., coverage fails) if and only if $V^{(n+1)}>Q_{1-\alpha}\left({\textstyle \sum_{i=1}^{n}}\frac{1}{n+1}\cdot\delta_{V^{(i)}}+\frac{1}{n+1}\cdot\delta_{+\infty}\right)$. According to Lemma A.1. in \cite{guan2023localized}, this is also equivalent to $V^{(n+1)}>Q_{1-\alpha}\left({\textstyle \sum_{i=1}^{n+1}}\frac{1}{n+1}\cdot\delta_{V^{(i)}}\right)$, that is $n+1\in S(V)$. Therefore, we have
\begin{align*}
	\mathbb{P}\left\{Y^{(n+1)}\notin\hat{\mathcal{C}}_{n}\left(X^{(n+1)},M^{(n+1)}\right)\right\}&=\mathbb{P}\{n+1\in S(V)\}=\frac{1}{n+1}\sum_{i=1}^{n+1}\mathbb{P}\{i\in S(V)\} \\
	&=\frac{1}{n+1}\mathbb{E}\left\{\sum_{i=1}^{n+1}\text{1}\left\{i\in S(V)\right\}\right\}=\frac{1}{n+1}\mathbb{E}\left\{|S(V)|\right\}\\
	&\leq\frac{1}{n+1}\cdot \alpha(n+1)=\alpha,
\end{align*}
where the second equality holds by the exchangeability of $V^{(i)}$, for $i\in [n+1]$. Finally,
\begin{equation*}
	\mathbb{P}\left\{Y^{(n+1)}\in\hat{\mathcal{C}}_{n}\left(X^{(n+1)},M^{(n+1)}\right)\right\}=1-\mathbb{P}\left\{Y^{(n+1)}\notin\hat{\mathcal{C}}_{n}\left(X^{(n+1)},M^{(n+1)}\right)\right\}\geq 1-\alpha,
\end{equation*}
which completes the proof.

\subsection{Proof of Lemma \ref{lemma:nonexchangeable conformal prediction}.}

The proof is similar to \cite{barber2023conformal}, where they consider cases without missingness. Recall that $R=(V^{(1)},\cdots,V^{(n+1)})$. And for any $K\in [n+1], R^{K}\in\mathbb{R}^{(n+1)}$ with entries
\begin{equation*}
	(R^{K})_{i} =
	\begin{cases}
		V^{(i)} & \text{if } i\neq K \text{ and } i\neq n+1, \\
		V^{(n+1)} & \text{if } i=K, \\
		V^{(K)} & \text{if } i=n+1.
	\end{cases}
\end{equation*}

The definition of the exchangeable conformal prediction in (\ref{nonexchangeable conformal prediction}) reveals 
\begin{equation}\label{proof:lemma1_1}
	Y^{(n+1}\notin\hat{\mathcal{C}}_{n}\left(X^{(n+1)},M^{(n+1)}\right)\Longleftrightarrow V^{(n+1)}> Q_{1-\alpha}\left(\sum_{i=1}^{n}\tilde{w}_{i}\cdot\delta_{V^{(i)}}+\tilde{w}_{n+1}\cdot\delta_{+\infty}\right).
\end{equation}
Next, we can verify that
\begin{equation}\label{proof:lemma1_2}
	Q_{1-\alpha}\left(\sum_{i=1}^{n}\tilde{w}_{i}\cdot\delta_{V^{(i)}}+\tilde{w}_{n+1}\cdot\delta_{+\infty}\right)\geq Q_{1-\alpha}\left(\sum_{i=1}^{n+1}\tilde{w}_{i}\cdot\delta_{(R^{K})_{i}}\right).
\end{equation}
If $K=n+1$, then $R^{K}=R$, and the bound holds trivially. If $K\leq n$, then the distribution on the left-hand side in (\ref{proof:lemma1_2}) can be written as
\begin{equation*}
	\sum_{i=1}^{n}\tilde{w}_{i}\cdot\delta_{V^{(i)}}+\tilde{w}_{n+1}\cdot\delta_{+\infty}=\sum_{i\in [n]:i\neq K} \tilde{w}_{i}\cdot\delta_{V^{(i)}}+\tilde{w}_{K}(\delta_{V^{(K)}}+\delta_{+\infty})+(\tilde{w}_{n+1}-\tilde{w}_{K})\delta_{+\infty},
\end{equation*}
while the distribution on the right-hand side in (\ref{proof:lemma1_2}) is 
\begin{align*}
	\sum_{i=1}^{n+1}\tilde{w}_{i}\cdot\delta_{(R^{K})_{i}}&=\sum_{i\in [n]:i\neq K} \tilde{w}_{i}\cdot\delta_{V^{(i)}}+\tilde{w}_{K}\cdot\delta_{V^{(n+1)}}+\tilde{w}_{n+1}\cdot\delta_{V^{(K)}} \\
	&=\sum_{i\in [n]:i\neq K} \tilde{w}_{i}\cdot\delta_{V^{(i)}}+\tilde{w}_{K}(\delta_{V^{(K)}}+\delta_{V^{(n+1)}})+(\tilde{w}_{n+1}-\tilde{w}_{K})\delta_{V^{(K)}}.
\end{align*}
Since $w_{i}=\rho^{n+1-i},i\in [n+1]$ are in increasing order, we have $\tilde{w}_{n+1}\geq\tilde{w}_{K}$, which verifies that (\ref{proof:lemma1_2}) must hold. Combining (\ref{proof:lemma1_1}) and (\ref{proof:lemma1_2}), we have
\begin{equation*}
	Y^{(n+1}\notin\hat{\mathcal{C}}_{n}\left(X^{(n+1)},M^{(n+1)}\right)\Longrightarrow V^{(n+1)}\geq Q_{1-\alpha}\left(\sum_{i=1}^{n+1}\tilde{w}_{i}\cdot\delta_{(R^{K})_{i}}\right),
\end{equation*}
equivalently,
\begin{equation}\label{proof:lemma1_3}
	Y^{(n+1}\notin\hat{\mathcal{C}}_{n}\left(X^{(n+1)},M^{(n+1)}\right)\Longrightarrow (R^{K})_{K}\geq Q_{1-\alpha}\left(\sum_{i=1}^{n+1}\tilde{w}_{i}\cdot\delta_{(R^{K})_{i}}\right).
\end{equation}

In the following, we define a function $S$ from $\mathbb{R}^{n+1}$ to subsets of $[n+1]$: for any $r\in\mathbb{R}^{n+1}$,
\begin{equation}\label{proof:lemma1_4}
	S(r)=\left\{i\in[n+1]:r_{i}>Q_{1-\alpha}\left(\sum_{j=1}^{n+1}\tilde{w}_{j}\cdot\delta_{r_{j}}\right)\right\}.
\end{equation}
According to Lemma A.1 in \cite{harrison2012conservative}, for any $r\in\mathbb{R}^{n+1}$,
\begin{equation}\label{proof:lemma1_5}
	\sum_{i\in S(r)}\tilde{w}_{i}\leq\alpha. 
\end{equation}
From (\ref{proof:lemma1_3}), it implies 
\begin{equation}\label{proof:lemma1_6}
	Y^{(n+1}\notin\hat{\mathcal{C}}_{n}\left(X^{(n+1)},M^{(n+1)}\right)\Longrightarrow K\in S(R^{K}).
\end{equation}
Then,
\begin{align*}
	\mathbb{P}\left\{K\in S(R^{K})\right\}&=\sum_{i=1}^{n+1}\mathbb{P}\left\{K=i\text{ and } i\in S(R^{i})\right\}\\
	&=\sum_{i=1}^{n+1}\tilde{w}_{i}\cdot\mathbb{P}\left\{i\in S(R^{i})\right\}\\
	&\leq\sum_{i=1}^{n+1}\tilde{w}_{i}\cdot\left(\mathbb{P}\left\{i\in S(R)\right\}+\text{d}_{\text{TV}}(R,R^{i})\right)\\
	&=\mathbb{E}\left\{\sum_{i\in S(R)}\tilde{w}_{i}\right\}+\sum_{i=1}^{n}\tilde{w}_{i}\cdot\text{d}_{\text{TV}}(R,R^{i})\\
	&\leq\alpha+\sum_{i=1}^{n}\tilde{w}_{i}\cdot\text{d}_{\text{TV}}(R,R^{i}),
\end{align*}
where the second equality holds because $K\perp \!\!\! \perp R$ and $R^{i}$ is a function of the data $R$, therefore, $K\perp \!\!\! \perp R^{i}$, and the last step holds by (\ref{proof:lemma1_5}). Finally,
\begin{align*}
	\mathbb{P}\left\{Y^{(n+1)}\in\hat{\mathcal{C}}_{n}\left(X^{(n+1)},M^{(n+1)}\right)\right\}&=1-\mathbb{P}\left\{Y^{(n+1)}\notin\hat{\mathcal{C}}_{n}\left(X^{(n+1)},M^{(n+1)}\right)\right\}\\
	&\geq 1-\mathbb{P}\left\{K\in S(R^{K})\right\}\\
	&\geq 1-\alpha-\sum_{i=1}^{n}\tilde{w}_{i}\cdot\text{d}_{\text{TV}}(R,R^{i}),
\end{align*}
which completes the proof.

\subsection{Proof of Theorem \ref{theorem:nonexchangeable conformal prediction}.}

The proof of the Theorem \ref{theorem:nonexchangeable conformal prediction} is straightforward using the fact in Lemma 1. The key difference between the marginal and mask-conditional probability is whether the mask is taken as random. Note that for the marginal probability, the mask is random. By i.i.d. observations, $(X^{(i)},M^{(i)},Y^{(i)}),i\in[n+1]$, are exchangeable, and $V^{(i)},i\in[n+1]$, are exchangeable. Therefore, $R$ and $R^{i}$, for $i\in [n]$ share the same distribution, and their total variance is zero. Thus, the marginal validity holds.

As for the mask-conditional probability, the mask is fixed. The data points become $(X^{(i)},m^{(i)},Y^{(i)}),i\in[n+1]$. Data from the same mask are also exchangeable. However, data from different masks may come from different distributions, see, e.g., Example \ref{example}. As a result, scores from different masks may not be exchangeable. For instance, $V^{(n+1)}=|Y^{(n+1)}-\hat{\mu}\circ\hat{\Phi}(X^{(n+1)},m^{(n+1)})|$, for $m^{(i)}\neq m^{(n+1)}$, the distribution of $(X^{(i)},Y^{(i)})$ may not be identical to that of $(X^{(n+1)},Y^{(n+1)})$; therefore, $(X^{(i)},Y^{(i)})$ and $(X^{(n+1)},Y^{(n+1)})$ may not be exchangeable. Thus, $V^{(i)}=|Y^{(i)}-\hat{\mu}\circ\hat{\Phi}(X^{(i)},m^{(i)})|$ may not be exchangeable to $V^{(n+1)}=|Y^{(n+1)}-\hat{\mu}\circ\hat{\Phi}(X^{(n+1)},m^{(n+1)})|$. Therefore, $R$ and $R^{i}$ may not be exchangeable, and their total variance may not be zero. 

\subsection{Proof of Theorem \ref{theorem:nexCP_new}.}
Note that the mask-conditional validity implies the marginal validity since marginalizing over $M$ for the mask-conditional probability leads to the marginal probability. Therefore, we will focus on the mask-conditional validity. The key point for mask-conditional validity is that the mask is taken as fixed, not random. 

Recall that if $(X,Y)$ is fully observed, the distribution of $(X,Y)$ can be written as $P_{(X,Y)}:=P_{Y|X}\cdot P_{X}$. When there exists a distribution shift, the distribution of $P_{Y|X}$ or $P_{X}$ may change. For example, when the distribution $P_{X}$ shifts to a different distribution, while the conditional distribution $P_{Y|X}$ remains unchanged, this phenomenon is referred to as covariate shift \citep{tibshirani2019conformal}. Similarly, in the setting of missing covariates, we can denote the ``observed" data as $(X_{\text{obs}(m)},m,Y)$. While $m$ is taken as fixed, not random, the ``observed" data can be rewritten as $(X_{\text{obs}(m)}Y)|M=m$, then its distribution can be express as 
\begin{equation}\label{appendix:joint distribution of test point}
	P_{(X_{\text{obs}(m)},Y)|M=m}=P_{Y|(X_{\text{obs}(m)},M=m)}\cdot P_{X_{\text{obs}(m)}|M=m}.
\end{equation}

In the proposed nonexchangeable conformal prediction (Algorithm \ref{algorithm:nexCP_new}), we find the available cases for the mask of the test point, i.e., $\tilde{\mathcal{I}}=\{i\in[n]:M^{(i)}\preceq M^{(n+1)}\}$. As we focus on the mask-conditional validity, in the subsequent analysis, we will set $M^{(n+1)}=m$. Note that for $\tilde{m}\preceq m$, the distribution of $(X_{\text{obs}(\tilde{m})},\tilde{m},Y)$ can be written as 
\begin{equation}\label{appendix:joint distribution of train point}
	P_{(X_{\text{obs}(\tilde{m})},Y)|M=\tilde{m}}=P_{Y|(X_{\text{obs}(\tilde{m})},M=\tilde{m})}\cdot P_{X_{\text{obs}(\tilde{m})}|M=\tilde{m}}
\end{equation}
By comparing equation (\ref{appendix:joint distribution of test point}) and (\ref{appendix:joint distribution of train point}), both the conditional distribution and marginal distribution change.

By masking the available data, $(X_{\text{obs}(\tilde{m})},\tilde{m},Y)$ changes to $(X_{\text{obs}(m)},\tilde{m},Y)$, and its distribution is 
\begin{equation}\label{appendix:train data by assumption}
	P_{(X_{\text{obs}(m)},Y)|M=\tilde{m}}=P_{Y|(X_{\text{obs}(m)},M=\tilde{m})}\cdot P_{X_{\text{obs}(m)}|M=\tilde{m}}.
\end{equation}
By comparing equation (\ref{appendix:joint distribution of test point}) and (\ref{appendix:train data by assumption}), $P_{Y|(X_{\text{obs}(m)},M=\tilde{m})}=P_{Y|(X_{\text{obs}(m)},M=m)}$ by Assmption (\ref{general assumption}), and $P_{X_{\text{obs}(m)}|M=\tilde{m}}=P_{X_{\text{obs}(m)}|M=m}$ by the MCAR Assumption. Thus, there is no distribution shift between $(X_{\text{obs}(m)},m,Y)$ and $(X_{\text{obs}(m)},\tilde{m},Y)$, for $\tilde{m}\preceq m$. Therefore, $V^{(n+1)}=|Y^{(n+1)}-\hat{\mu}\circ\hat{\Phi}(X^{(n+1)},m)|$ and $\tilde{V}^{(i)}=|Y^{(i)}-\hat{\mu}\circ\hat{\Phi}(X^{(i)},m)|,i\in\tilde{\mathcal{I}}$ are exchangeable, so $\tilde{V}^{(i)}_{*},i\in[|\tilde{\mathcal{I}}|]$ and $V^{(n+1)}$ are exchangeable. There is no coverage loss by introducing fixed weights when the scores are indeed exchangeable \citep{barber2023conformal}. Thus, for the fixed weights $\tilde{w}_{i},i\in[|\tilde{\mathcal{I}}|$ and $\tilde{w}_{n+1}$, the nonexchangeable conformal prediction in Algorithm \ref{algorithm:nexCP_new} satisfyes
\begin{align*}
	\mathbb{P}\left\{Y^{(n+1)}\in\hat{\mathcal{C}}_{n}\left(X^{(n+1)},m\right)|M=m\right\}&=\mathbb{P}\left\{V^{(n+1)}\leq Q_{1-\alpha}\left({\textstyle \sum_{i\in [|\tilde{\mathcal{I}}|]}}\tilde{w}_{i}\cdot\delta_{\tilde{V}_{*}^{(i)}}+\tilde{w}_{n+1}\cdot\delta_{+\infty}\right)\right\}\\
	&\geq 1-\alpha,
\end{align*}
which completes the proof.

\subsection{Proof of Theorem \ref{theorem:LCP}.}

In the first part of the proof, we show the mask-conditional validity of the localized conformal prediction. Consequently, marginalizing over $M$ leads to the marginal validity. In the second part of the proof, we show the asymptotically conditional validity.

{\noindent\bf Mask-condtional validity.} Note that 
\begin{align*}
	&\mathbb{P}\left\{Y^{(n+1)}\in\hat{\mathcal{C}}_{n}\left(X^{(n+1)},m\right)\mid M^{(n+1)}=m\right\}\\
	&=\mathbb{P}\left\{V_{\alpha,h}\left(X^{(n+1)},m,Y^{(n+1)}\right)\leq Q_{1-\alpha}\left(\frac{1}{|\tilde{\mathcal{I}}|+1}\sum_{i\in\tilde{\mathcal{I}}}\delta_{V_{\alpha,h}^{(i)}}+\frac{1}{|\tilde{\mathcal{I}}|+1}\delta_{+\infty}\right)\right\},
\end{align*}
where $V_{\alpha,h}^{(i)}=\tilde{V}^{(i)}-Q_{1-\alpha}\left(\hat{F}_{h}^{(i)}\right),\tilde{V}^{(i)}=|Y^{(i)}-\hat{\mu}\circ\hat{\Phi}(X^{(i)},m)|$, and $\hat{F}_{h}^{(i)}={\textstyle \sum_{i\in\tilde{\mathcal{I}}}}p_{h}(X^{(i)}|x)\text{1}(v\leq\tilde{V}^{(i)})$, for $i\in\tilde{\mathcal{I}}$. Note that the mask is fixed. 

Define a monotonic score function $s(v,F,\theta)=Q_{1-\alpha}(F)+\theta-v$ with respect to $\theta$, and the inverse function of $s$ as $p(v,F,t)={\textstyle \inf_{\theta\in\mathbb{R}}}\left\{\theta:s(v,F,\theta)\geq t\right\}$. We know that $p(v,F,0)=v-Q_{1-\alpha}(F)$. For simplification, let $\hat{F}^{(i)}=\hat{F}^{(i)}_{h}(v|X=X^{(i)},M=m),\tilde{V}^{(i)}=V(X^{(i)},m,Y^{(i)}),\theta^{(i)}=p(\tilde{V}^{(i)},\hat{F}^{(i)},0), i\in\tilde{\mathcal{I}}$, and
\begin{equation*}
	\theta^{*}=Q_{1-\alpha}\left(\frac{1}{|\tilde{\mathcal{I}}|+1}\sum_{i\in[\tilde{\mathcal{I}}]}\delta_{\theta^{(i)}}+\frac{1}{|\tilde{\mathcal{I}}|+1}\delta_{+\infty}\right).
\end{equation*}
Suppose that $E_{\tilde{v}}$ is the event when $\{\tilde{V}^{(i)},i\in[\tilde{\mathcal{I}}]\}=\{\tilde{v}^{(i)},i\in[\tilde{\mathcal{I}}]\}$. For a fixed $\theta$, let $S^{\theta}_{i}=s(\tilde{V}^{(i)},\hat{F}^{(i)},\theta)$ and $s^{\theta}_{i}$ as its realization when $\tilde{V}^{(i)}=\tilde{v}^{(i)}$. By quantile lemma, we have 
\begin{equation*}
	\mathbb{P}\left\{S^{\theta}_{n+1}\leq Q_{1-\alpha}\left(\frac{1}{|\tilde{\mathcal{I}}|+1}\sum_{i\in[\tilde{\mathcal{I}}]}\delta_{s^{\theta}_{i}}+\frac{1}{|\tilde{\mathcal{I}}|+1}\delta_{+\infty}\right)\mid E_{\tilde{v}},\theta\right\}\geq 1-\alpha.
\end{equation*}
Denote 
\begin{equation*}
	h(\theta)=Q_{1-\alpha}\left(\frac{1}{|\tilde{\mathcal{I}}|+1}\sum_{i\in[\tilde{\mathcal{I}}]}\delta_{s_{i}^{\theta}}+\frac{1}{|\tilde{\mathcal{I}}|+1}\delta_{+\infty}\right).
\end{equation*}
By the definition of $\theta^{(i)}$, we have $s_{i}^{\theta^{(i)}}=0$. In addition, by the monotonicity of $s(v,F,\theta)$ with respect to $\theta$ and the definitional of $\theta^{*}$, that is, $\theta^{*}$ is less or equal to at least the $(1-\alpha)$-quantile of all $\theta^{(i)}$ and $\{+\infty\}$, we have at least $(1-\alpha)$-quantile of $s_{i}^{\theta^{*}}\leq 0$. So, one get $h(\theta^{*})\leq 0$. Therefore, for the fixed $\theta^{*}$ given ecent $E_{\tilde{v}}$, we have
\begin{equation*}
	\mathbb{P}\left\{S_{n+1}^{\theta^{*}}\leq h(\theta^{*})\leq 0|E_{\tilde{v}}\right\}\geq 1-\alpha.
\end{equation*}
Marginalizing for event possible event, we have
\begin{equation*}
	\mathbb{P}\left\{s(\tilde{V}_{n+1},\hat{F}^{(n+1)},\theta^{*})\leq 0\right\}\geq 1-\alpha.
\end{equation*}

Letting $\hat{\mathcal{C}}_{n}(X^{(n+1)},m)=\left\{y\in\mathbb{R}:s(\tilde{V}^{(n+1)},\hat{F}^{(n+1)},\theta^{*})\leq 0\right\}$, where $\tilde{V}^{(n+1)}=|y-\hat{\mu}\circ\hat{\Phi}(X^{(n+1)},m)|$, and $\hat{F}^{(n+1)}$ is the approximate CDF of $F(v|X=X^{(n+1)},M=m)$, we have 
\begin{equation*}
	\mathbb{P}\left\{Y^{(n+1)}\in\hat{\mathcal{C}}_{n}(X^{(n+1)},m)|M^{(n+1)}=m\right\}\geq 1-\alpha.
\end{equation*}
We know that the $\hat{\mathcal{C}}_{n}(X^{(n+1)},m)$ is the proposed localized conformal prediction, which completes the proof. 

{\noindent\bf Asymptotically conditional validity.}

Denote $\hat{p}_{h}(x)$ as the empirical distribution estimated by $K(d(X^{(i)}.x)/h)$, $p(x)$ as the original distribution, that is, $\hat{p}_{h}(x)=K(d(X^{(i)},x)/h)/h^{k}$, and $\mathbb{E}[\hat{p}_{h}(x)]=p_{h}(x)$, where $k$ is the dimension of $X$. By proposition A.1 of \cite{han2022split}, when $h=O(n^{-1/(2\gamma+k)})$, under Assumption \ref{holder_clsss_condition},
\begin{equation*}
	\sup_{p\in\Sigma(\gamma,L)}\mathbb{E}\left[\int(\hat{p}_{h}(x)-p_{h}(x))^{2}dx \right]=O\left(\left(\frac{1}{n}\right)^{\frac{2\gamma}{2\gamma+k}}\right),
\end{equation*}
and 
\begin{equation*}
	\sum_{p\in\Sigma(\gamma,L)}Var[\hat{p}_{h}(x)]=O\left(\left(\frac{1}{n}\right)^{\frac{2\gamma}{2\gamma+k}}\right).
\end{equation*}
Therefore, 
\begin{equation*}
	\hat{p}_{h}(x)=p_{h}(x)+O_{p}\left(\left(\frac{1}{n}\right)^{\frac{\gamma}{2\gamma+k}}\right)
\end{equation*}
over $x$ in its support. Plug into the approximation of condition density function $\hat{F}_{h}$, one gets
\begin{align*}
	\mathbb{E}\int\left[\hat{F}_{h}(V\mid X=x,M=m)-F_{h}(V|X=x,M=m) \right]^{2}dx \\
	\leq\sup_{p\in\Sigma(\gamma,L)}\mathbb{E}\int(\hat{p}_{h}(x)-p(x))^{2}dx=O\left(\left(\frac{1}{n}\right)^{\frac{2\gamma}{2\gamma+k}}\right),
\end{align*}
which implies that when $n\to\infty$, that is, if a sufficient amount of data is available, our nanparametric estimation of conditional distribution $\hat{F}_{h}(V\mid X=x,M=m)$ will converge to the true conditional distribution $F(V\mid X=x,M=m)$. Therefore, the localized conformal prediction with provide a new score function $V_{\alpha,h}\approx \tilde{V}(x,m,y)-q_{v,\alpha,m}(x)\approx V(x,m,y)-q_{v,\alpha,m}(x)$, where $q_{v,\alpha,m}(x)$ is the true quantile function $q_{v,\alpha,m}(x)=\inf\{v:F(v|X=x,M=m)\geq\alpha\}$. So, $\mathbb{P}\{V_{\alpha,h}\geq 0\mid X=x,M=m\}\approx\alpha$ is uniform for all $x$ and $m$. The marginal quantile correction of $V_{\alpha,h}$ will close to 0 and given an asymptotic conditonal coverage guarantee for the split localized conformal in Algorithm \ref{algorithm:nexCP_new},
\begin{equation*}
	\mathbb{P}\{Y^{(n+1)}\in\hat{\mathcal{C}}_{n}(x,m)\mid X^{(n+1)}=x, M^{(n+1)}=m\}\geq 1-\alpha,
\end{equation*}
as $n\to\infty$.


\bibliography{reference}

\begin{thebibliography}{48}
\providecommand{\natexlab}[1]{#1}
\providecommand{\url}[1]{{#1}}
\providecommand{\urlprefix}{URL }
\providecommand{\doi}[1]{\url{https://doi.org/#1}}
\providecommand{\eprint}[2][]{\url{#2}}
 \bibcommenthead

\bibitem[{Angelopoulos et~al.(2023)Angelopoulos, Bates
  et~al.}]{angelopoulos2023conformal}
Angelopoulos AN, Bates S, et~al (2023) Conformal prediction: A gentle
  introduction. Foundations and Trends{\textregistered} in Machine Learning
  16(4):494--591

\bibitem[{Angelopoulos et~al.(2024)Angelopoulos, Bates, Fisch, Lei, and
  Schuster}]{angelopoulos2024conformal}
Angelopoulos AN, Bates S, Fisch A, et~al (2024) Conformal risk control. In: The
  Twelfth International Conference on Learning Representations,
  \urlprefix\url{https://openreview.net/forum?id=33XGfHLtZg}

\bibitem[{Ayme et~al.(2022)Ayme, Boyer, Dieuleveut, and Scornet}]{ayme2022near}
Ayme A, Boyer C, Dieuleveut A, et~al (2022) Near-optimal rate of consistency
  for linear models with missing values. In: International Conference on
  Machine Learning, PMLR, pp 1211--1243

\bibitem[{Ayme et~al.(2023)Ayme, Boyer, Dieuleveut, and
  Scornet}]{ayme2023naive}
Ayme A, Boyer C, Dieuleveut A, et~al (2023) Naive imputation implicitly
  regularizes high-dimensional linear models. In: International Conference on
  Machine Learning, PMLR, pp 1320--1340

\bibitem[{Bao et~al.(2024)Bao, Huo, Ren, and Zou}]{bao2024selective}
Bao Y, Huo Y, Ren H, et~al (2024) Selective conformal inference with false
  coverage-statement rate control. Biometrika 111(3):727--742

\bibitem[{Barber et~al.(2023)Barber, Candes, Ramdas, and
  Tibshirani}]{barber2023conformal}
Barber RF, Candes EJ, Ramdas A, et~al (2023) Conformal prediction beyond
  exchangeability. The Annals of Statistics 51(2):816--845

\bibitem[{Bates et~al.(2023)Bates, Cand{\`e}s, Lei, Romano, and
  Sesia}]{bates2023testing}
Bates S, Cand{\`e}s E, Lei L, et~al (2023) Testing for outliers with conformal
  p-values. The Annals of Statistics 51(1):149--178

\bibitem[{Berrett and Samworth(2023)}]{berrett2023optimal}
Berrett TB, Samworth RJ (2023) Optimal nonparametric testing of missing
  completely at random and its connections to compatibility. The Annals of
  Statistics 51(5):2170--2193

\bibitem[{Cand{\`e}s et~al.(2023)Cand{\`e}s, Lei, and
  Ren}]{candes2023conformalized}
Cand{\`e}s E, Lei L, Ren Z (2023) Conformalized survival analysis. Journal of
  the Royal Statistical Society Series B: Statistical Methodology 85(1):24--45

\bibitem[{Cannings and Fan(2022)}]{cannings2022correlation}
Cannings TI, Fan Y (2022) The correlation-assisted missing data estimator.
  Journal of Machine Learning Research 23(41):1--49

\bibitem[{Dua et~al.(2017)Dua, Graff et~al.}]{dua2017uci}
Dua D, Graff C, et~al (2017) Uci machine learning repository

\bibitem[{Follain et~al.(2022)Follain, Wang, and Samworth}]{follain2022high}
Follain B, Wang T, Samworth RJ (2022) High-dimensional changepoint estimation
  with heterogeneous missingness. Journal of the Royal Statistical Society
  Series B: Statistical Methodology 84(3):1023--1055

\bibitem[{Foygel~Barber et~al.(2021)Foygel~Barber, Candes, Ramdas, and
  Tibshirani}]{barber2021limits}
Foygel~Barber R, Candes EJ, Ramdas A, et~al (2021) The limits of
  distribution-free conditional predictive inference. Information and
  Inference: A Journal of the IMA 10(2):455--482

\bibitem[{Guan(2023)}]{guan2023localized}
Guan L (2023) Localized conformal prediction: A generalized inference framework
  for conformal prediction. Biometrika 110(1):33--50

\bibitem[{Gui et~al.(2023)Gui, Barber, and Ma}]{gui2023conformalized}
Gui Y, Barber R, Ma C (2023) Conformalized matrix completion. Advances in
  Neural Information Processing Systems 36:4820--4844

\bibitem[{Han et~al.(2022)Han, Tang, Ghosh, and Liu}]{han2022split}
Han X, Tang Z, Ghosh J, et~al (2022) Split localized conformal prediction.
  arXiv preprint arXiv:220613092

\bibitem[{Harrison(2012)}]{harrison2012conservative}
Harrison MT (2012) Conservative hypothesis tests and confidence intervals using
  importance sampling. Biometrika 99(1):57--69

\bibitem[{Hore and Barber(2024)}]{hore2024conformal}
Hore R, Barber RF (2024) Conformal prediction with local weights: randomization
  enables robust guarantees. Journal of the Royal Statistical Society Series B:
  Statistical Methodology p qkae103

\bibitem[{Josse and Reiter(2018)}]{josse2018introduction}
Josse J, Reiter JP (2018) Introduction to the special section on missing data.
  Statistical Science 33(2):139--141

\bibitem[{Josse et~al.(2024)Josse, Chen, Prost, Varoquaux, and
  Scornet}]{josse2024consistency}
Josse J, Chen JM, Prost N, et~al (2024) On the consistency of supervised
  learning with missing values. Statistical Papers pp 1--33

\bibitem[{Le~Morvan et~al.(2020{\natexlab{a}})Le~Morvan, Josse, Moreau,
  Scornet, and Varoquaux}]{le2020neumiss}
Le~Morvan M, Josse J, Moreau T, et~al (2020{\natexlab{a}}) Neumiss networks:
  differentiable programming for supervised learning with missing values.
  Advances in Neural Information Processing Systems 33:5980--5990

\bibitem[{Le~Morvan et~al.(2020{\natexlab{b}})Le~Morvan, Prost, Josse, Scornet,
  and Varoquaux}]{le2020linear}
Le~Morvan M, Prost N, Josse J, et~al (2020{\natexlab{b}}) Linear predictor on
  linearly-generated data with missing values: non consistency and solutions.
  In: International Conference on Artificial Intelligence and Statistics, PMLR,
  pp 3165--3174

\bibitem[{Le~Morvan et~al.(2021)Le~Morvan, Josse, Scornet, and
  Varoquaux}]{le2021sa}
Le~Morvan M, Josse J, Scornet E, et~al (2021) What’s a good imputation to
  predict with missing values? Advances in Neural Information Processing
  Systems 34:11530--11540

\bibitem[{Lee et~al.(2024)Lee, Dobriban, and Tchetgen}]{lee2024simultaneous}
Lee Y, Dobriban E, Tchetgen ET (2024) Simultaneous conformal prediction of
  missing outcomes with propensity score $\epsilon$-discretization. arXiv
  preprint arXiv:240304613

\bibitem[{Lei and Wasserman(2014)}]{lei2014distribution}
Lei J, Wasserman L (2014) Distribution-free prediction bands for non-parametric
  regression. Journal of the Royal Statistical Society Series B: Statistical
  Methodology 76(1):71--96

\bibitem[{Lei et~al.(2018)Lei, G’Sell, Rinaldo, Tibshirani, and
  Wasserman}]{lei2018distribution}
Lei J, G’Sell M, Rinaldo A, et~al (2018) Distribution-free predictive
  inference for regression. Journal of the American Statistical Association
  113(523):1094--1111

\bibitem[{Little(1993)}]{little1993pattern}
Little RJ (1993) Pattern-mixture models for multivariate incomplete data.
  Journal of the American Statistical Association 88(421):125--134

\bibitem[{Little and Rubin(2019)}]{little2019statistical}
Little RJ, Rubin DB (2019) Statistical analysis with missing data, vol 793.
  John Wiley \& Sons

\bibitem[{Liu et~al.(2024)Liu, Zhan, and Lin}]{liu2024penalized}
Liu Z, Zhan Z, Lin C (2024) Penalized regression analysis with
  individual-specific patterns of missing covariates. Communications in
  Statistics-Simulation and Computation 53(7):3126--3142

\bibitem[{Mao et~al.(2024)Mao, Martin, and Reich}]{mao2024valid}
Mao H, Martin R, Reich BJ (2024) Valid model-free spatial prediction. Journal
  of the American Statistical Association 119(546):904--914

\bibitem[{Pedregosa et~al.(2011)Pedregosa, Varoquaux, Gramfort, Michel,
  Thirion, Grisel, Blondel, Prettenhofer, Weiss, Dubourg
  et~al.}]{pedregosa2011scikit}
Pedregosa F, Varoquaux G, Gramfort A, et~al (2011) Scikit-learn: Machine
  learning in python. Journal of Machine Learning Research 12:2825--2830

\bibitem[{Romano et~al.(2019)Romano, Patterson, and
  Candes}]{romano2019conformalized}
Romano Y, Patterson E, Candes E (2019) Conformalized quantile regression.
  Advances in neural information processing systems 32

\bibitem[{Rubin(1976)}]{rubin1976inference}
Rubin DB (1976) Inference and missing data. Biometrika 63(3):581--592

\bibitem[{Santos et~al.(2020)Santos, Abreu, Wilk, and
  Santos}]{santos2020distance}
Santos MS, Abreu PH, Wilk S, et~al (2020) How distance metrics influence
  missing data imputation with k-nearest neighbours. Pattern Recognition
  Letters 136:111--119

\bibitem[{Seedat et~al.(2023)Seedat, Jeffares, Imrie, and van~der
  Schaar}]{seedat2023improving}
Seedat N, Jeffares A, Imrie F, et~al (2023) Improving adaptive conformal
  prediction using self-supervised learning. In: International Conference on
  Artificial Intelligence and Statistics, PMLR, pp 10160--10177

\bibitem[{Sell et~al.(2024)Sell, Berrett, and Cannings}]{sell2024nonparametric}
Sell T, Berrett TB, Cannings TI (2024) Nonparametric classification with
  missing data. The Annals of Statistics 52(3):1178--1200

\bibitem[{Sesia and Romano(2021)}]{sesia2021conformal}
Sesia M, Romano Y (2021) Conformal prediction using conditional histograms.
  Advances in Neural Information Processing Systems 34:6304--6315

\bibitem[{Shafer and Vovk(2008)}]{shafer2008tutorial}
Shafer G, Vovk V (2008) A tutorial on conformal prediction. Journal of Machine
  Learning Research 9(3)

\bibitem[{Shao and Zhang(2023)}]{shao2023distribution}
Shao M, Zhang Y (2023) Distribution-free matrix prediction under arbitrary
  missing pattern. arXiv preprint arXiv:230511640

\bibitem[{Tibshirani et~al.(2019)Tibshirani, Foygel~Barber, Candes, and
  Ramdas}]{tibshirani2019conformal}
Tibshirani RJ, Foygel~Barber R, Candes E, et~al (2019) Conformal prediction
  under covariate shift. Advances in neural information processing systems 32

\bibitem[{Vovk et~al.(2005)Vovk, Gammerman, and Shafer}]{vovk2005algorithmic}
Vovk V, Gammerman A, Shafer G (2005) Algorithmic learning in a random world,
  vol~29. Springer

\bibitem[{Wang(2015)}]{wang2015tying}
Wang Y (2015) Tying the Autocrat's hands. Cambridge University Press

\bibitem[{Wilson and Martinez(1997)}]{wilson1997improved}
Wilson DR, Martinez TR (1997) Improved heterogeneous distance functions.
  Journal of artificial intelligence research 6:1--34

\bibitem[{Yin et~al.(2024)Yin, Shi, Wang, and Blei}]{yin2024conformal}
Yin M, Shi C, Wang Y, et~al (2024) Conformal sensitivity analysis for
  individual treatment effects. Journal of the American Statistical Association
  119(545):122--135

\bibitem[{Zaffran et~al.(2023)Zaffran, Dieuleveut, Josse, and
  Romano}]{zaffran2023conformal}
Zaffran M, Dieuleveut A, Josse J, et~al (2023) Conformal prediction with
  missing values. In: International Conference on Machine Learning, PMLR, pp
  40578--40604

\bibitem[{Zaffran et~al.(2024)Zaffran, Josse, Romano, and
  Dieuleveut}]{zaffran2024predictive}
Zaffran M, Josse J, Romano Y, et~al (2024) Predictive uncertainty
  quantification with missing covariates. arXiv preprint arXiv:240515641

\bibitem[{Zhan et~al.(2023)Zhan, Li, and Zhang}]{zhan2023partial}
Zhan Z, Li X, Zhang J (2023) Partial replacement imputation estimation for
  partially linear models with complex missing pattern covariates. Statistics
  and Computing 33(4):88

\bibitem[{Zhu et~al.(2022)Zhu, Wang, and Samworth}]{zhu2022high}
Zhu Z, Wang T, Samworth RJ (2022) High-dimensional principal component analysis
  with heterogeneous missingness. Journal of the Royal Statistical Society
  Series B: Statistical Methodology 84(5):2000--2031

\end{thebibliography}

\end{document}